\documentclass[physics,article,submit,unicode-math,pdftex,moreauthors]{Definitions/mdpi}

\def\mbh{$M_{\rm BH}$\/}
\def\nh{$n_{\mathrm{H}}$\/}
\def\lledd{$L/L_{\rm Edd}$}

\def\nc{$N_{\rm c}$\/}
\def\rfe{$R_{\rm FeII}$}

\def\feiiq{\rm Fe{\sc ii}$\lambda$4570\/}
\def\msol{M$_\odot$\/}

\def\ltsima{$\; \buildrel < \over \sim \;$}
\def\ltsim{\lower.5ex\hbox{\ltsima}} 
\def\gtsima{$\; \buildrel > \over \sim \;$}

\def\gtsim{\lower.5ex\hbox{\gtsima}} 

\def\lya{{ Ly}$\alpha$}
\def\civfull{{\sc{Civ}}$\lambda$1549\/}
\def\civ{{\sc{Civ}}\/}

\def\cm3{cm$^{-3}$\/}
\def\hb{{\sc{H}}$\beta$\/}

\def\mgii{{Mg\sc{ii}}$\lambda$2800\/}

\def\oiiiopt{{\sc{[Oiii]}}$\lambda\lambda$4959,5007\/}

\def\siiii{Si{\sc iii]}\/}
\def\aliii{Al{\sc iii}\/}
\def\ciii{C{\sc iii]}\/}

\def\heiiuv{He{\sc{ii}}$\lambda$1640}
\def\nv{{N{\sc v}}\/}

\def\feii{{Fe\sc{ii}}\/}
\def\siii{{Si\sc{ii}}$\lambda$1814\/}

\def\fe{{\sc{Fe}}\/}

\def\heiiopt{He{\sc{ii}}$\lambda$4686\/}
\def\fe76087{{\sc [Fe vii]}$\lambda$6087\/}

\def\kms{km~s$^{-1}$}

\def\ergss{erg\, s$^{-1}$\/}
\def\hii{H{\sc ii}\/}

\def\heiiopt{{\sc H}e{\sc ii}$\lambda$4686\/}

\def\siiv{Si{\sc iv}}
\def\oiv{O{\sc iv]}$\lambda$1402\/}

\usepackage{enumerate}
\usepackage{ulem}
\definecolor{darkorange}{rgb}{1,0.612,0}
\definecolor{aquamarine}{rgb}{0.498,1,0.8314}

\defcitealias{marzianietal03a}{M03}
\defcitealias{marzianietal09}{M09}
\defcitealias{sulenticetal07}{S07}
\defcitealias{zamfiretal10}{Z10}
\makeatletter
\DeclareRobustCommand{\onontimes}{%
 \mathbin{\mathpalette\on@ntimes\relax}%
}
\newcommand{\on@ntimes}[2]{%
 \vcenter{\hbox{%
 \sbox0{\m@th$#1\otimes$}%
 \setlength\unitlength{\wd0}%
 \begin{picture}(1,1)
 \linethickness{0.35pt}
 \put(.5,.5){\circle{.8}}
 \end{picture}%
 }}%
}
\makeatother
\firstpage{1} 
\pubvolume{1}
\issuenum{1}
\articlenumber{0}
\pubyear{2023}
\copyrightyear{2023}
\datereceived{ } 
\daterevised{ } % Comment out if no revised date
\dateaccepted{ } 
\datepublished{ } 
\hreflink{https://doi.org/} % If needed use \linebreak
\nolinenumbers
\Title{From sub-solar to super-solar chemical abundances along the quasar main sequence}

\TitleCitation{Metallicity along the quasar main sequence}

\Author{Paola Marziani$^{1,*}$\orcidA{}, Alberto Floris$^{2}$, Alice Deconto-Machado$^{3}$\orcidD, Swayamtrupta Panda$^{4}$\thanks{CNPq Fellow}\orcidB{}, Marzena Sniegowska$^{5}$, Karla Garnica$^{6}$, Deborah Dultzin$^{6}$, Mauro D'Onofrio$^{2,1}$, Ascensi\'on Del Olmo$^{3}$\orcidC{}, Edi Bon$^{7}$, and Nata\v{s}a Bon$^{7}$}%, and Who Else$^{?}$} 

\longauthorlist{yes}

\address{%
$^{1}$ National Institute for Astrophysics (INAF), Astronomical Observatory of Padova, IT-35122 Padova, Italy\\
$^{2}$ Dipartimento di Fisica e Astronomia, Universit\`a di Padova, Vicolo dell' Osservatorio 3, IT-35122 Padova, Italy\\
$^{3}$ Instituto de Astrof\'\i sica de Andaluc\'\i a\ (IAA-CSIC), Glorieta de Astronom\'\i a s/n, 38038 Granada, Spain \\
$^{4}$ Laborat\'{o}rio Nacional de Astrofísica, MCTI, R. dos Estados Unidos 154, Na\c{c}\~{o}es, CEP 37504-364, Itajub\'a, Brazil\\
$^{5}$ School of Physics and Astronomy, Tel Aviv University, Tel Aviv, 69978, Israel\\
$^{6}$ Instituto de Astronom\'{\i}a, UNAM, Mexico D.F. 04510, Mexico\\ 
$^{7}$ Belgrade Astronomical Observatory, Volgina 7, Belgrade, Serbia\\
}

\corres{Correspondence: paola.marziani@inaf.it}

\abstract{The 4D Eigenvector 1 sequence has proven to be a highly effective tool for organizing observational and physical properties of type 1 active galactic nuclei (AGN). In this paper, we present multiple measurements of metallicity for the broad line region gas, from new or previously published data. We demonstrate a consistent trend along the optical plane of the E1 (also known as the quasar main sequence), defined by the line width of \hb\ and by a parameter measuring the prominence of singly-ionized iron emission. The trend involves an increase from sub-solar metallicity in correspondence with extreme Population B (weak \feii\ emission, large \hb\ FWHM) to metallicity several tens the solar value in correspondence with extreme Population A (very strong \feii\ optical emission, narrower \hb\ profiles). The data establish the metallicity as a correlate of the 4D E1/main sequence. If the very high metallicity ($Z \gtrsim 10 Z_\odot$) gas is expelled from the sphere of influence of the central black hole, as indicated by the widespread evidence of nuclear outflows and disk wind in the case of sources radiating at high Eddington ratio, then it is possible that the outflows from quasars played a role in chemically enriching the host galaxy.
}


\begin{document}

%%%%%%%%%%%%%%%%%%%%%%%%%%%%%%%%%%%%%%%%%%
\section{Introduction: A Main Sequence and the Eigenvector 1 for Quasars}

The main sequence (MS) concept in quasar research draws parallels with stellar evolutionary sequences, but in this case, it is applied to quasar properties. A study on a sample of Palomar-Green quasars detected an interesting anti-correlation between the strength of the \feiiq\ emission line and the Full Width at Half Maximum (FWHM) {of the broad \hb{} emission line} of type-1 quasars with low redshifts ($z < 1$, \cite{borosongreen92}). This anti-correlation suggests that, as the \feiiq\ feature becomes stronger, the FWHM tends to decrease, and is one of the main correlations known as the Eigenvector 1 of quasars. This finding has been established through the analysis of samples of increasing size over the years \cite{gaskell85,sulenticetal00c,zamfiretal10,shenho14}, and has proved to be fundamental to organize type-1 AGN properties in a systematic way with predictive ability. 
%\section{Spectral Types and Multifrequency Diversity}

Quasars are categorized into different spectral types along the MS \cite[e.g.,][]{shenho14}, and two primary populations, referred to as Population A and Population B, have been identified \cite{sulenticetal00c}. These categories are based on specific properties such as the Eddington ratio (a measure of the accretion rate) and orientation \cite{shenho14,sunshen15, pandaetal19}. The "4DE1" classification, introduced by \citet{sulenticetal00c}, further organizes quasar properties according to the Eddington ratio and orientation, revealing a systematic pattern of variation across different types of AGN encompassing their outflow phenomenology and their accretion mode \cite{giustiniproga19}. The assignation of most quasar spectral types permitted the prediction of their UV, X-ray, radio, and FIR properties with a high degree of confidence. 

%\section{The Main Sequence as an Evolutionary Scheme}
The MS, much like the equivalent concept in stellar evolution, is used as an evolutionary framework for understanding quasars. This sequence spans from young and rejuvenated quasars, characterized by specific spectral properties (Extreme Population A), to older and more mature quasars with distinct characteristics (Population B). Differences in factors such as black hole mass, Eddington ratio, disk winds, or outflow properties, derived from the strengths and profiles of emission lines are used to define and distinguish these evolutionary stages \cite{duetal16,fraix-burnetetal17}.

{ Currently, the available evidence regarding the correlation between metal content in the broad line emitting region (BLR) and the quasar main sequence (MS) remains partial and inconclusive \cite{pandaetal19, Panda_2021}.}
%At present, the evidence is partial and inconclusive on whether the metal content in the broad line emitting region (BLR) is a correlate of the quasar MS \cite{pandaetal19}. 
There is a long tradition of studies attempting to estimate the metallicity in the BLR of AGN over a broad range of redshift, from $z \approx 0$, up to $z \approx 6$ \cite[e.g.,][]{hamannferland93,nagaoetal06,juarezetal09,matsuokaetal11,shinetal13,sameshimaetal17,wangetal22}. All these studies derive metallicity in the range from a few times solar to about 10 times solar, values that are significantly higher than the ones found even in most massive and metal-rich galaxies \cite{matteucci12,xuetal18}. When considering the past metallicity estimates along the main sequence, an intriguing trend emerges: not all quasars are accreting matter with super-solar metal content \cite{marzianietal23}, and only at one end of the MS, the BLR gas may be enriched by a metal content even above ten times solar, possibly with pollution by supernova ejecta \cite{sniegowskaetal21}. 

In this contribution, we will use the results of a new analysis and of several recent papers to gather a view of the global trend along the MS. Section \ref{obs} summarizes the new observations from ground- and space-based observatories. Section \ref{methods} elucidates the methodology applied to the metallicity estimates, stressing the need to isolate line components that are spectroscopically resolved, and may correspond to emitting regions in different physical conditions (\S \ref{multi}). Results of several individual sources and composite spectra representative of entire spectral types (ST) introduced in \S \ref{cases} clearly delineate a sequence of increasing metallicity from extreme Population B to extreme Population A (\S \ref{results}). As outflows from the AGN are of special relevance for galactic evolution \citep{dimatteoetal05,hopkinsetal06,somervilledave15}, it is important to isolate the corresponding line component, whenever possible (\S \ref{blue}). The AGN outflows may provide enrichment of the nuclear and circumnuclear region of the host galaxies (\S \ref{disc}), although ensuing chemical feedback is expected to be relevant only at high AGN luminosity. In \S\ \ref{disc}, we outline how the different metallicities inferred for the BLR of low-$z$ AGN fit the evolutionary interpretation of the quasar MS.

\section{Observations}
\label{obs}
The new spectral data employed in this study were acquired through a series of distinct observations. Specifically:

\begin{itemize}
 \item For Mrk 335, Mrk 478, and Fairall 9, we utilized optical spectra sourced from \cite{marzianietal03a}. Additionally, the UV spectra of Mrk 335 were obtained during observations conducted on the 4th and 7th of January 2013, utilizing the Cosmic Origins Spectrograph (COS) aboard the Hubble Space Telescope (HST) with the G140L grism. The UV spectra of Mrk 478 were acquired on the 5th of December 1996, utilizing the HST's Faint Object Spectrograph (FOS) and employing the G130H and G190H grisms. For Fairall 9, UV spectra were collected on the 22nd of January 1993, utilizing the HST's FOS with the G190H and G270H grisms, and subsequently on the 18th of July 2012, employing the HST's COS with the G130M and G160M grisms.
 \item For NGC 1275, the data were sourced from \cite{punslyetal18}. Optical spectra encompassed various observations spanning the period from 1983 to 2017. Additionally, UV spectra were  { acquired from the HST MAST}, with FOS observations in 1993 and COS observations in 2011. 
 \item For PHL 1092, data were drawn from \cite{marinelloetal20b}. Optical spectra were captured using the Goodman spectrograph at the 4.1-meter telescope of the Southern Observatory for Astrophysical Research (SOAR) on the night of 12th December 2014. The UV spectra were obtained using the HST's Space Telescope Imaging Spectrograph (STIS) on the night of 20th August 2003.
\end{itemize}

In addition, we considered: 
\begin{itemize}
 \item composite spectra for radio-loud (RL) and radio-quiet (RQ) Population B sources. The data on which the composites were based were described in a previous work \cite{marzianietal23};
\item median results for two samples of xA sources at intermediate redshift ($z \approx 2$) \cite{sniegowskaetal21,garnicaetal22} lacking the rest-frame optical spectrum. 
\end{itemize}

%Composite spectra(?)

%\swayam{\textbf{Paola, the bib file, in its current format, can't be edited. Probably it is too big. Perhaps we can shift the useful references to a new file where we can edit/add references.}} 

\section{Methodology and Data analysis}
\label{methods}

\subsection{Multicomponent analysis}
\label{multi}
A crucial technique in quasar studies is the multicomponent analysis, which involves dissecting complex emission line profiles into distinct components. This approach allows us to uncover details about different regions of emission, potentially revealing information about kinematics, velocities, and structures. Line ratios, such as those involving \civ$\lambda$1549, \heiiuv, \aliii$\lambda 1860$, and \siiv+\oiv, are used as diagnostic tools to infer physical conditions, such as ionization and metallicity, within the quasar environments. In the spectra of quasars, both low-ionization and high-ionization optical and UV lines are observed (LILs and HILs, respectively), and they provide valuable insights into the physical conditions and structure of the quasar BLR. These lines are emitted, at least in part, in distinct regions with different characteristics \citep{collinsouffrinetal88,elvis00}. 

We distinguish a broad component (BC) that is broadened by the Doppler effect due to the rapid motion of the gas in the BLR in a velocity field dominated by Keplerian motions \cite{petersonwandel99,petersonwandel00}. Lines meeting this condition are also referred to as “virialized broad emission lines." Low-ionization lines like \hb\ and \mgii\ are typically emitted in the virialized BLR, which may be characterized by high densities and column densities, but relatively low ionization (ionization parameter $\sim 10^{-2}$. High-ionization lines like \civ$\lambda1549$, \heiiuv\, and \nv$\lambda1240$\ are also emitted in the virialized BLR. However, these lines trace the regions of the BLR that are exposed to the most intense and energetic radiation from the accretion disk \cite{collinsouffrinetal88,elvis00}. They are often broader than low-ionization lines due to higher velocities in this part of the BLR, and show prominent blueshifts \cite{richardsetal11,coatmanetal16}. 

In short, we subdivide all lines into three main components that can account for the diversity of line profiles, along the quasar MS, resulting from the balance between gravitation and radiation forces \cite{ferlandetal09}. 

\subsubsection{Population A}

\begin{description}
 \item[Broad Component (BC)] Represented by a Lorentzian function, symmetric and unshifted or slightly redshifted \citep{veroncettyetal01,zamfiretal10}. 
 \item[Blueshifted Component (BLUE)] It is defined as the excess of emission on the blue side of the BC. The shape can be irregular, but the profiles resemble "triangular" or "trapezoidal shapes" \citep{marzianietal17c,vietrietal18} that are usually well modeled by asymmetric Gaussian \cite{marzianietal17c,deconto-machadoetal23}. The blueshifted component can be very prominent at high Eddington ratios and high luminosity, dominating the HIL emission \citep{leighlymoore04,mejia-restrepoetal16,vietrietal20}. The BLUE is increasing in prominence in the HILs along the quasar main sequence and reaches its maximum at ST A3-A4, where \rfe\ is also at a maximum. 
\end{description}

The two components most likely represent coexisting regions \cite{wangetal11}, albeit in very different physical and dynamical conditions. While the BC is associated with a dense, low ionization region, capable of emitting mainly (but not exclusively) LILs and maintaining a virialized velocity field (for which a large column density is needed \cite{marconietal09}), the outflowing gas should be of higher ionization. The assumption that BLUE and BC refer to regions with the same metallicity has been questioned by models of the AGN involving nuclear star formation \cite{wangetal06}, and in the following, we will attempt independent metallicity estimates from BLUE and BC.
%but seems to be a reasonable assumption. 

\subsubsection{Population B}

\begin{description}
 \item[Broad Component (BC)] Represented by a Gaussian function, symmetric and unshifted or slightly redshifted \citep{zamfiretal10}.
 \item[Very Broad Component (VBC)] Represented by a Gaussian function, redshifted by about $\sim 2000$\kms \citep{zamfiretal10, wolfetal20}. Given the virial velocity field of the emitting regions, this component represents the innermost emission of the BLR. Several works have described the Population B Balmer profiles in terms of a BLR and a very broad line region (VBLR) \cite{petersonferland86,sneddengaskell04}. It is unclear whether the emitting gas might be so highly ionized to be optically thin to the Lyman continuum \cite{morrisward89}. The origin of the redshift is the subject of current debate, and two main alternatives have been proposed: gravitational redshift \cite{netzer77,corbin95,popovicetal95,munozetal03,mediavillaetal18,fianetal22}, and infall \cite{wangetal14b}. The data unambiguously support the gravitational redshift hypothesis only for $\log$\mbh $\gtrsim 8.7$ [\msol], while lower \mbh\ require very low \lledd\ for the profiles to show a significant gravitational effect \cite{marziani23a}. 
 \item[Blueshifted Component (BLUE)] It is defined as the excess of emission on the blue side of the BC+VBC profile. The blueshifted component is usually not prominent at low Eddington ratios but can still affect the centroid and asymmetry index of both HILs and LILs. {Due to its weakness,} BLUE is always modeled by a shifted (symmetric) Gaussian.
\end{description}

\begin{figure}[t!]
\centering
\includegraphics[width=10.5 cm]{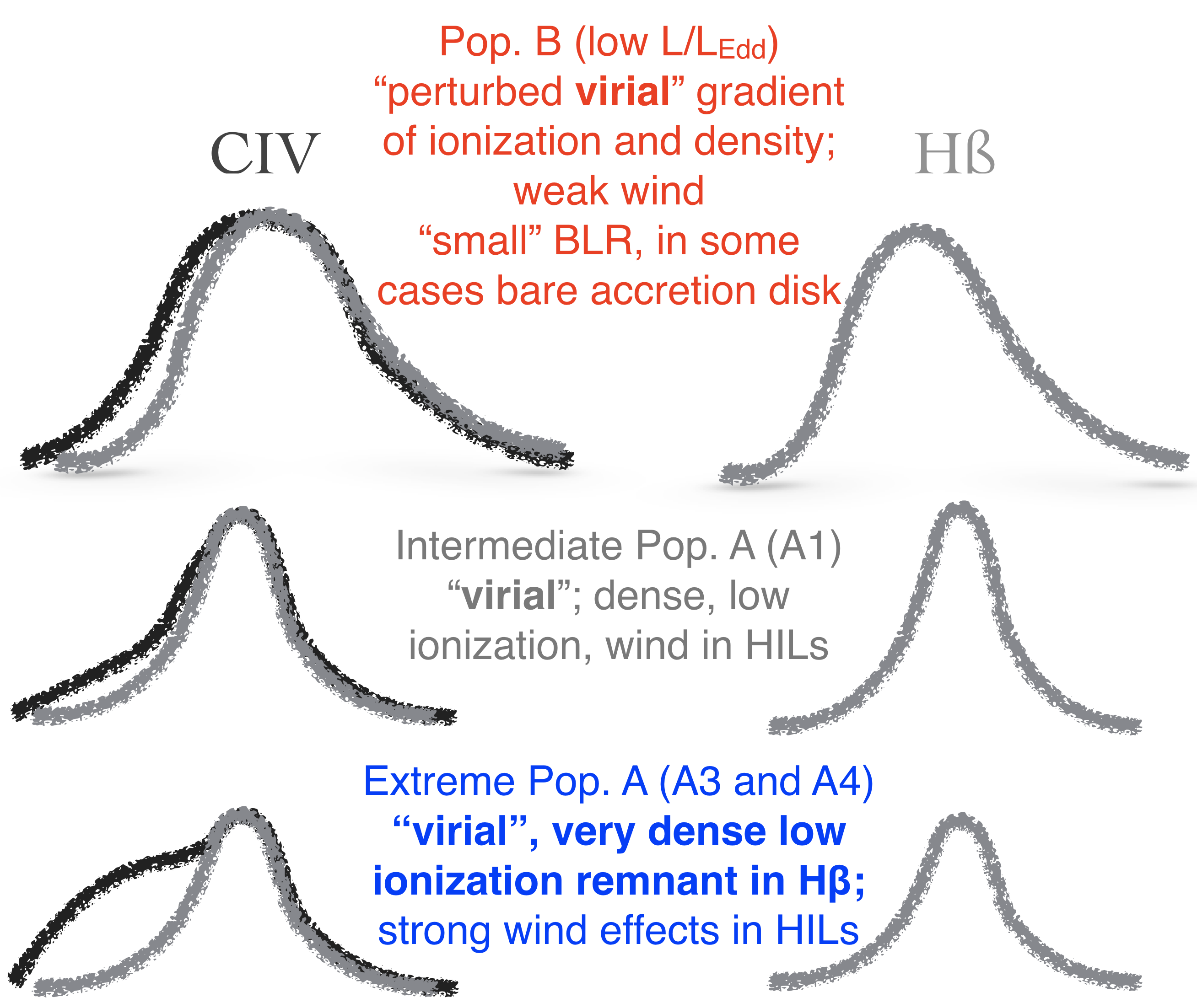}
\caption{Interpretation of the line profiles of low- and high-ionization lines along the MS for isolating major spectroscopically resolved components. \label{fig:interp}}
\end{figure} 
\unskip
Fig. \ref{fig:interp} summarizes the interpretation of the line profiles of \civ\ and \hb\ assumed as prototypical HIL and LIL, respectively, for Population B, Population A, and extreme Population A. Since BLUE is barely resolved in Pop. B, no estimates of the metallicity $Z$ will be attempted. For the object of extreme Population A an estimate has been carried out for BC and BLUE, while for Pop. B sources, for the BC and VBC. In the latter case, we expect that the metallicity is the same, but the physical conditions would reflect a gradient in ionization (lower ionization for the BC and higher for the VBC), and hence be different on average. 

\subsection{Emission line ratios}

%Generally speaking 

\begin{description}
\item[Z indicator \civ/\siiv+\oiv] has been widely applied as metallicity indicator \citep[e.g.,][]{nagaoetal06,shinetal13}. In photoionization equilibrium, the classical argument derived for \hii\ regions that the electron temperature decreases with increasing metallicity \cite{pageletal79} works for the BLR as well. The intensity of the \civ\ line actually decreases as the metal abundance increases. However, the reason why the ratio \civ/\siiv\ is a metallicity indicator resides in the "competition" of He$^+$\ ions that have roughly the same creation potential of C$^{++}$. As a result, the Str\"omgren sphere of C$^{+3}$\ decreases much more strongly with increasing $Z$\ than for Si$^{+3}$: the ionization potential of Si$^{+2}$\ is 2.46 Ryd, so the relatively unabsorbed continuum between 2.5 and 3.5 Ryd is available to maintain
a proportionality with its abundance \cite{huangetal23}. This effect dominates over the lower electron temperature that affects the collisional excitation rates of both \siiv\ and \civ, expected to be higher for \siiv. 

\item[Z indicators \civ/\heiiuv\ and \siiv/\-\heiiuv] When considering indicators like \civ/\-\heiiuv\ and \siiv/\-\heiiuv, they should show sensitivity to the abundance of Carbon and Silicon, assuming that the ratio of Helium relative to Hydrogen remains constant. However, the dependence of \civ/\heiiuv\ on $Z$\ is not monotonic: it increases for sub-solar metallicities and then declines steadily up to 200$Z_{\odot}$, for specific conditions with $\log U \sim 0$\ and $\log$\nh $\sim 9$\cm3\ \cite{sniegowskaetal21}. For lower $U$ values, the behavior is monotonic \cite{garnicaetal22}. This underscores the necessity for multiple intensity ratios that depend on $Z$\ and $U$. 
\item[$Z$\ indicators involving \nv, \nv/\civ\ and \nv/\heiiuv] have also been extensively employed in previous studies \cite{hamannferland93,ferlandetal96,nagaoetal06,hamannetal02,shinetal13}. The strength of the \nv\ line was unexpectedly high in a photoionization scenario, possibly due to a selective enhancement of nitrogen \citep[e.g.,][]{osmersmith76,shields76}, { resulting from} secondary production of N by massive and intermediate-mass stars, and yielding [N/H]$\propto Z^{2}$\ \citep{vila-costasedmunds93,izotovthuan99,matsuokaetal11}. This process may be particularly significant in cases of abnormal star formation and evolution processes that are expected to occur within active nuclei. Contamination by narrow and semi-broad absorption features is often significant, and even with precise modeling of high-ionization lines, it may be challenging to reconstruct the unabsorbed profile of the red wing of Ly$\alpha$. In this analysis, we refrain from using ratios involving \nv\ because they are not consistently measurable. 
%In the cases where they were measured, they are added a posteriori and discussed to verify the consistency of the metallicity estimates with and without considering \nv.
\item[Density\ indicators] The ratios \aliii/\-\siiii\ and \siiii/\-\ciii\ { are} responsive to density since they involve intercombination lines with well-defined critical densities ($n_\mathrm{c}\sim 10^{10}$ cm$^{-3}$ for \ciii\ \citep{hamannetal02} and $n_\mathrm{c}\sim$ 10$^{11}$ cm$^{-3}$ for \siiii). 
\item[Ionization parameter] \siiii/\siiv, \siii/\siiii, and \siii/\siiv\ are influenced by the ionization parameter and remain insensitive to changes in $Z$\ since the lines are from different ionic states of the same element. Also, the ratio \ciii/\civ\ is sensitive to the ionization parameter but entails a strong dependence on the \nh\ as well. The ratio \civ/\hb\ is also a clear diagnostic, although it is also dependent on $Z$\ and, unfortunately, often made unreliable by the intrinsic variations of the quasar and by poor photometric accuracy if observations are not synoptic and dedicated. 
\item[Mixed diagnostics: \feii/\hb\ ] The ratio \rfe\ deserves a particular attention. As with any other metal to Hydrogen ratio, it entails an obvious dependence on iron abundance and hence on metallicity. Nonetheless, \rfe\ is dependent on density, ionization parameter and column density of the line emitting gas, in the sense that large \rfe\ ($\gtrsim 1$) seem possible only for relatively high \nh\ ($\gtrsim 10^{11}$ \cm3), low ionization and large \nc\ ($\gtrsim 10^{23}$ cm$^{-2}$) \cite{collinjoly00,matsuokaetal08,pandaetal19}. 
\end{description} 

In the following, we will try to use the same ratios as much as possible for the three-line components. {However,} the BLUE components are often { so} weak to be undetectable in several LIL profiles. In this case, we consider upper or lower limits as appropriate. Table \ref{tab:ratios} provides an overview of the measured intensity ratios applied to the metallicity estimates of several of our targets. 

\begin{table}[htp]
\fontsize{7pt}{7pt}\selectfont
\caption{Diagnostic intensity ratios \label{tab:diags}}\hspace{-5cm}
\begin{center}\setlength{\tabcolsep}{2pt}
\begin{tabular}{lcccccccccccc}
\hline
Component & Si{\sc iv+Oiv]}/ &Si{\sc iv+Oiv]}/ &C{\sc iv}/ & C{\sc iv}/ &C{\sc iv}/ &Al{\sc iii}/ & Al{\sc iii}/ & Al{\sc iii}/ & Si{\sc iii]}/ &\rfe & He{\sc ii}$_\mathrm{opt}$ \\
 & C{\sc iv} & He {\sc ii}$_\mathrm{UV}$ & He {\sc ii}$_\mathrm{UV}$ & C{\sc iii}] & \hb\ & C{\sc iv}& He {\sc ii}$_\mathrm{UV}$ & Si{\sc iii}] & C{\sc iii}] & & \hb \\ \hline
BLUE [A] & \textcolor{blue}{${{\onontimes}}$} & \textcolor{blue}{$\onontimes$} & \textcolor{blue}{$\onontimes$} & {---} & $>$ & $<$ & $<$ & --- & ---- & $<$ & --- \\
BC & \textcolor{blue}{$\onontimes$} & \textcolor{blue}{$\onontimes$} & \textcolor{blue}{$\onontimes$} & \textcolor{blue}{$\onontimes$} & \textcolor{cyan}{$\onontimes$} & \textcolor{blue}{$\onontimes$} & {$\onontimes$} & \textcolor{blue}{$\onontimes$}&\textcolor{blue}{$\onontimes$} & \textcolor{blue}{$\onontimes$} & $<$ \\
VBC [B] & \textcolor{blue}{$\onontimes$} & \textcolor{blue}{$\onontimes$} & \textcolor{blue}{$\onontimes$} & \textcolor{blue}{$\onontimes$} & \textcolor{cyan}{$\onontimes$} & {$\onontimes$} & --- & $<$ & $<$ & $\onontimes$ & $\onontimes$ \\
 \hline
\end{tabular}
\end{center}
{ $\onontimes$ measured ratio with associated uncertainty. $<$\ and $>$: upper and lower limit to intensity ratio, respectively. ---: not available. He {\sc ii}$_\mathrm{UV}$\ and He {\sc ii}$_\mathrm{opt}$ are \heiiuv\ and He {\sc ii}$\lambda$4686, respectively. Blue circles identify the ratios actually used for the new sources presented in this work; the use of the \civ/\hb\ ratio has been considered only for Fairall 9, due to the non-contemporaneity of the rest-frame optical and UV data. Unavailable ratios involve two undetectable components.} 
\label{tab:ratios}
\end{table}%

%FWHM(\hb) and HST/FOS-based Composites:
%The FWHM of the \hb emission line serves as an important parameter for characterizing quasar properties. By studying composite spectra obtained using the Hubble Space Telescope's Faint Object Spectrograph (HST/FOS), we combined data from a sample of quasars to gain insights into common features: the composites are however computed only for the sources belonging spectral type. These composites cover various aspects, including the \civ\ and \hb\ lines, which are analyzed to identify trends and characteristics shared among these distant objects.

\subsection{Photoionization simulations}

Understanding the physical conditions within quasar environments involves estimating parameters like metallicity ($Z$), density (\nh), and ionization state (ionization parameter $U$). The three fundamental parameters can be estimated by comparing observed line ratios with model predictions obtained through computational models, such as CLOUDY, which simulate the interactions between radiation and gas in the environments of the broad line region \cite{ferlandetal17}. Input parameters for photoionization computations are the photoionizing continuum spectral energy distribution (SED), the ionization parameter (or an alternative, luminosity, and distance of the emitting region), gas hydrogen density, chemical composition, and a micro-turbulence parameter. There is evidence of trends for all of these parameters along the quasar main sequence, and in the case of Population B, there is evidence of a radial stratification of the properties within the BLR \cite[][]{baldwinetal95,koristaetal97} that is heuristically modeled separating a BC and VBC (BLR and VBLR). 

The arrays of simulations were therefore organized as follows: 5 different SEDs, one for each of the following cases: Pop. B RL, Pop. B RQ, with a dedicated SED for NGC 1275, an SED for Pop. A sources \cite{mathewsferland87} and one for extreme Pop. A \citep[high Eddington ratio of Ref.][]{ferlandetal20}. Metallicity ($Z$) was assumed to scale as solar ($Z_\odot$), with 12 values ranging between $0.01$\ and 1000 $Z_\odot$ for Pop. A and 14 values between $0.001$\ and 20 $Z_\odot$ for Pop. B. The micro-turbulence parameter was set to 0 \kms. This is relatively insignificant for resonance UV lines \cite{sniegowskaetal21}, but is expected to lead to an under-prediction of \feii\ emission \cite{sigutpradhan98,sigutpradhan03,sniegowskaetal21}. For each metallicity value, we considered an array of simulations covering the \nh\ and $U$\ parameter plane in the range $7 \le \log $\nh$\le 14$ \cm3, $-4.5 \le \log U \le 1$ for Pop. A (667 simulations), and $7 \le \log $\nh$\le 13$ \cm3, $-3 \le \log U \le 1$\ for Pop. B (425 simulations). For each source or composite spectra, the set of $\approx 8 - 9$\ diagnostic ratios were compared with a set of $\approx$ 8000 and 6000 simulations covering the parameter space \nh, $U$, and $Z$, for Population A and B respectively. The computations were carried out independently for the three components identified in the emission lines, as they are thought to represent distinct regions in different physical conditions. The solution for the single zone model (i.e., a single point in the 3D space \nh, $U$, $Z$) was identified by the minimum $\chi^2$\ computed from the difference between the observed line ratios and the predicted line ratios over the entire 3D space \citep[see, e.g.,][]{sniegowskaetal21,garnicaetal22}. 

Errors on measured line ratios were estimated assuming that the continuum placement was the dominant source of uncertainties and setting extreme continua as $cont \pm rms$, where $cont$\ is the best-fit continuum and $rms$\ is the noise measured over the continuum itself and propagated according to the ``triangular distribution" \cite{dagostini03}. Limits at 1 $\sigma$ and 90\%\ confidence were set by computing the ratio $F = \chi^2/\chi^2_\mathrm{min}$ between $\chi^2$\ of different models for $n_\mathrm{ratios} -3 $\ degrees of freedom. 

There are several caveats in the method, related to both the quality of the data and the model assumption: (1) the non-simultaneity of the observations in the optical and UV. Often, optical and UV observations are separated by years in the rest frame of the quasar. In addition, photometric inter-calibration between optical and UV data is problematic: while space-based observations are precise within a few percent, optical data are affected by uncontrolled light loss. As a result, the ratio \civ/\hb\ was measured but ultimately removed from the computations in the new sources analyzed in the present paper (Mark 335, Mark 478, PHL 1092) {except for} Fairall 9. In this case, simulations both with and without the ratio \civ/\hb\ were run and gave consistent results. (2) A major assumption is that metallicity scales as solar. Albeit this is a time-honored assumption \cite{nagaoetal06b,maiolinomannucci19}, it is not a reasonable one because major differences are expected for a disk star in a late-type spiral galaxy and the nuclear region of an active galaxy \cite[][and references therein]{wangetal23}. Actually, recent works discussed the evidence of pollution by supernova ejecta \cite{cantielloetal21,garnicaetal22}. (3) Photoionization computations are carried out under the assumption of single-zone emission. While this assumption seems a good one for extreme Population A, where density and ionization tend toward limiting values \cite{pandaetal19}, this might not be the case for Pop. B sources.

\section{Case studies}
\label{cases}
Basic properties of the cases considered in this paper are reported in Table \ref{tab:objects}, and include the E1 optical parameter FWHM \hb\ and \rfe, as well as the accretion parameters (luminosity, black hole mass \mbh, and Eddington ratio). The last columns provide the radio loudness parameter and some notes that list bibliographical sources of information or notable, recent work related to the specific object. 

\begin{description}
\item{Composite RL - ST B1} — A RL composite spectrum was obtained from 20 RL sources, with redshift range $\approx$0.25 -- 0.65, absolute magnitudes between $-$23.5 and $-$26.5 and~S/N $\approx$130 and $\approx$55 for the visual and UV ranges. The $Z$ estimates are used as provided by \citet{marzianietal23} since the method of analysis and measurement is basically the same. 
\item{NGC 1275 — } NGC 1275 (Perseus A) is an elliptical galaxy, and the brightest cluster galaxy of the Perseus Cluster, one of the most massive galaxy clusters in the nearby Universe. NGC 1275 is associated with a cooling flow phenomenon, where gas in the cluster's intracluster medium is thought to cool and sink towards the central regions of the galaxy \cite{heckmanetal89,limetal08}, potentially fueling its modest AGN activity. The photoionizing continuum, which is crucial for understanding the ionization state of the gas in NGC 1275, was defined {\em ad hoc} by observational constraints on the SED \citep[see][ for details]{punslyetal18}. The estimated photoionizing continuum had a spectral index $\alpha_{\nu}$ = 0.5 in the far-UV range (500 to 800 \AA) and a spectral break at 800 \AA. Beyond this spectral break, the spectral index $\alpha_{\nu}$ = 2 down to the soft X-ray band at 0.5 keV, with $\alpha_{\nu}$ = 1.0 in the harder X-ray band. In summary, the SED of the far-UV shown by \citet{punslyetal18} indicates a broad line Seyfert-like AGN with a soft ionizing continuum, a weak hard ionizing continuum, and no Compton deflection hump. The big blue bump associated with thermal emission from optically thick, geometrically thin accretion \cite{sunmalkan89} is weaker than the ones of the SED templates appropriate for Population B quasars, see Fig. 5 of \citet{punslyetal18}. The \lledd\ is extremely low and, interestingly, the BLR is correspondingly weak, to the point that a careful, dedicated analysis was needed to disentangle the broad line profiles from the much stronger narrow line emission. %\swayam{we may show a figure with all the SEDs considered in this work.}
\item{Composite RQ - ST B1 — } 
%Spectral data of good quality covering the full spectral ranges from 1000\,\AA\ to $\approx$6000\,\AA\ (i.e., from~\lya\ to \hb\ included) for the same object are not easily available for low-$z$ quasars, as~the UV coverage demands space-based observations.
A composite spectrum was constructed for a group of 16 RQ sources, all falling within the spectral bin B1, within the redshift range of approximately {redshift range $\approx $0.002--0.5}.These sources span an absolute magnitude range $-$21 -- $-$27, which corresponds to a bolometric luminosity range $\log L \sim$ 45--47\mbox{ [erg s$^{-1}$]}. To achieve comprehensive UV coverage from 1000 \AA\ to around 6000 Å (including the spectral region from \lya\ to \hb), which is typically demanding and necessitates space-based observations, the UV data were obtained from HST/FOS observations as discussed in \citet{sulenticetal07}. Additionally, the optical spectra were sourced from \cite{marzianietal03a}. The composite spectrum has a high signal-to-noise ratio (S/N) of approximately 90. The metallicities (Z values) for this composite and their corresponding uncertainties were adopted from \cite{marzianietal23}.

\item {Fairall 9 - ST B1 ---} Empirical parameters derived from the Main Sequence (MS) analysis indicate Fairall 9's spectral type as B1, a category well-populated among quasars along the MS. This classification consistently aligns with a low \lledd\ ratio. Notably, Fairall 9 exhibits radio quietness, as it eluded detection in the Sydney University Molonglo Sky Survey \cite{mauchetal03} with a detection limit of 6 mJy, implying a radio-to-optical specific flux ratio of $\lesssim$ 1.5. We {use} $\log R_{\rm K} \approx -0.04$\ \cite{sikoraetal07}. 

 The most recent assessment of the black hole mass includes estimates using the reverberation mapping technique that have converged on values ranging from (1.5 -- 2.5) $\times 10^8$ \msol\ \cite{petersonetal04,bentzkatz15}, contingent on the adopted virial factor, as well as a spectropolarimetric-derived \mbh\ that allowed for a virial factor estimate, yields $(1.5\pm 0.5) \cdot 10^8$ \msol. 
A conventional estimate of Fairall 9's bolometric luminosity stands at $\log L_\mathrm{bol} \approx$ 45.3 \ergss, with an Eddington ratio of $\log$ \lledd $\approx$ -2.0, placing it toward the lower end within the distribution of Population B sources. The Spectral Energy Distribution (SED) also conforms to the characteristics of Population B objects, devoid of a prominent big-blue bump. More details are given in a recent paper by \citet{jiangetal21}.
\item {Mrk 335 - ST A1 ---} Markarian 335 is a population A Seyfert 1, spectral type A1. It is located in the nearby Universe with a redshift of 0.0256. This AGN exhibits characteristics typical of an RQ A1 AGN, with lower-than-average \feii\ emissions, positioning it in the lower-left corner of the MS. Emission lines in the UV and optical ranges exhibit little to no blueshifts in their profiles. The only exception to this typical behavior is the positive slope of its optical continuum, which is likely due to galactic extinction.
\item {Mrk 478 - ST A2 ---} Markarian 478 is a Pop. A Seyfert 1, spectral type A2, borderline A3 { from the measurements of the present analysis}. It is located at a redshift of 0.077. Although classified as A2, it exhibits characteristics that suggest it may be an extreme accretor. Table \ref{tab:objects} shows that it has an Eddington ratio $\approx 1$. Additionally, it displays a strong \feii\ emission and a pronounced outflowing component in its emission line profiles. Hence, it could be argued that Mrk 478 could be classified as an A3-type object.
\item {PHL 1092 - ST A4 ---} Palomar Haro Luyten 1092 is a population A Seyfert 1, spectral type A4, located at a redshift of 0.3965. Its spectrum is characterized by strong UV emissions, with a notably sloped SED. PHL 1092 is considered an extremely accreting quasar, as it exhibits prime characteristics of one, including a strong outflowing component in its emission line profiles, particularly noticeable in \civ$\lambda1549$ and \siiv+\oiv, which casts a shadow on the virialized component. In the optical range, the \hb\ emission is overshadowed by the \feii\ emission, mirroring the \feii\ profile of I Zw 1, itself considered an extreme accretor.

\item{Extreme Population A (A3, A4) ---} The intermediate redshift xA sample of \citet{sniegowskaetal21} and \citet{garnicaetal22} allowed for $Z$ estimates from the UV spectral lines. The sample lacks the optical data providing the important information from the \hb\ spectral range, and the line width reported in Table \ref{tab:objects} is the one of \aliii. The sources were selected based on UV criteria that were found equivalent to the criterion \rfe $\gtrsim $ 1 for the identification of extreme Population A sources, at least at a high degree of confidence \cite{buendia-riosetal23}. Their luminosity is significantly higher than the luminosity of the other sources considered in this paper, although the Eddington ratio is consistent with the ones of Mark 478 and PHL 1092, $\sim \mathcal{O}(1)$.

\end{description}

\begin{figure}[t!]
\includegraphics[width=14. cm]{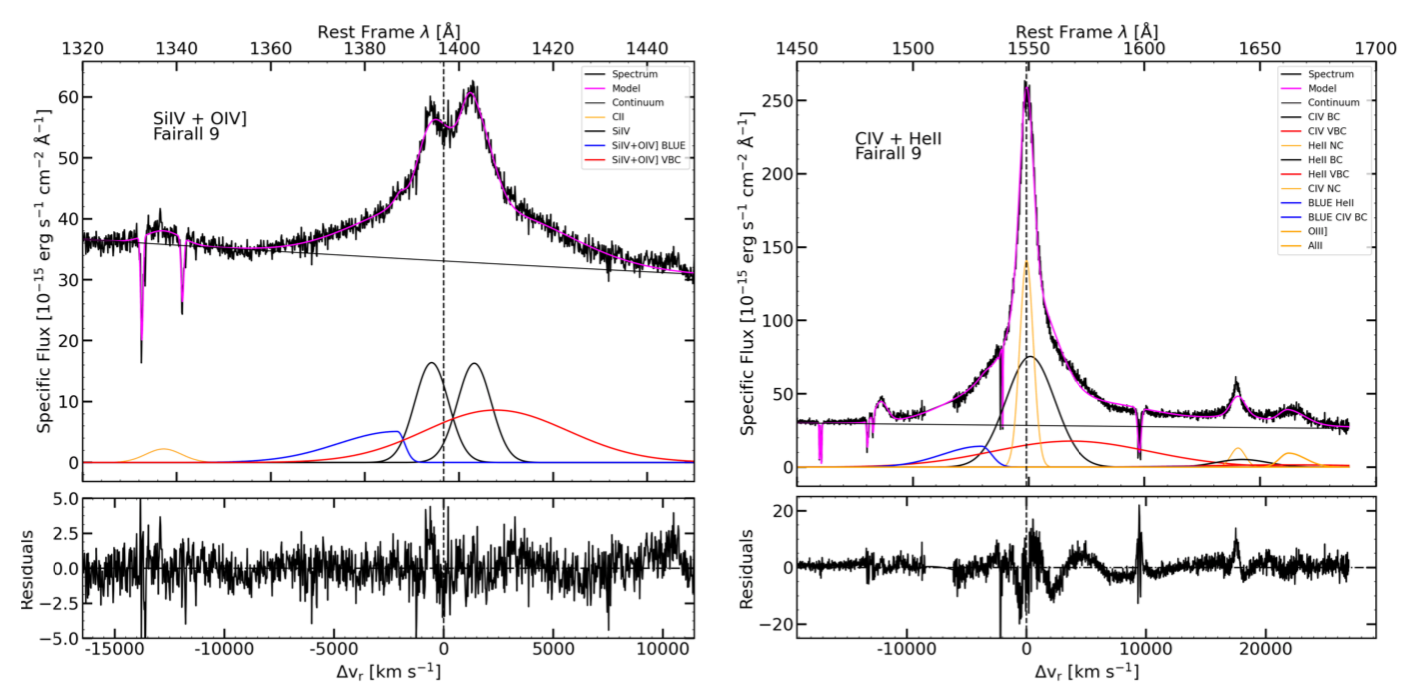}
\includegraphics[width=14. cm]{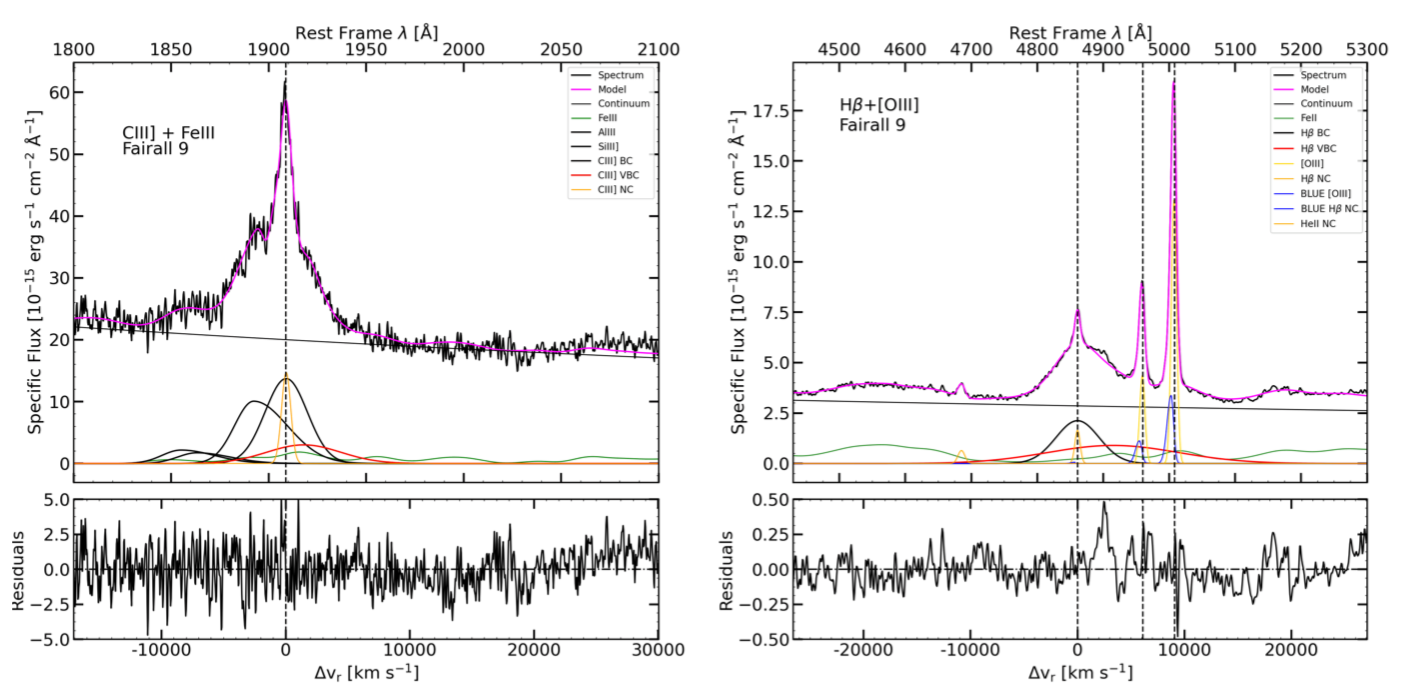}
\caption{\footnotesize Fits of the emission line spectrum of Fairall 9, the prototypical Population B source. From left to right, top row: 1400 \AA\ blend; \civ\ + \heiiuv\ blend; bottom row: 1900 \AA\ blend; \hb\ + \feii\ blend. The line profile of \hb\ and the 1900 blend are accounted for by two components, BC (black lines) and VBC (red). The HILs in the rest-frame ultraviolet show an excess with respect to the BC+VBC extending on the blue side of the line, modeled with a skew Gaussian \cite{azzaliniregoli12} and ascribed to emission from a spectroscopically resolved outflow that adds up to the virialized components. The magenta lines trace the full empirical model of the line profiles. The bottom panels show the (observed minus computed) residuals.\label{fig:f9fits}}
\end{figure} 
\unskip

\begin{table}[h!]
\fontsize{6pt}{6pt}\selectfont
\caption{Object properties}
\tabcolsep=2pt
\label{tab:objects} % is used to refer this table in the text
\centering % used for centering table
\begin{tabular}{c c c c c c c c c l} % centered columns (4 columns)
\hline\hline % inserts double horizontal lines
Object name & ST & $z$ & FWHM(H$\beta$) & \rfe \ & $L_\mathrm{bol}$ & M$_{\rm BH}$ & \lledd & $R_\mathrm{K}$ & Notes\\ 
& & & [km s$^{-1}$] & &[erg s$^{-1}$] & [M$_\odot$] & & \\
(1) & (2) & (3) & (4) & (5) & (6) & (7) & (8) & (9) & (10)\\
\hline 
Composite RL & B1 & $\approx$ 0.25-0.65 & 7690 &0.17 & $8\cdot 10^{44} - 8\cdot 10^{45}$ & $10^8 - 10^9$ & 0.01 -- 0.1 & $\gtrsim 80$ & \cite{marzianietal23} \\
NGC 1275 & B1 & 0.017 & 4770 & n.d.$^a$,$\lesssim 0.2$ &$1.2 \times 10^{43}$ & $8 \times 10^{8}$ & $\lesssim 10^{-4}$ & 1744 & \lledd, \mbh, $L$s: \cite{punslyetal18} \\
Composite RQ & B1 & $\approx$ 0.002-0.5 & 5530 & 0.34 &$8\cdot 10^{43} - 2\cdot 10^{46}$ & 4$\times 10^8$ -- $10^9$ & 0.01 - 0.04 & $\lesssim 10$ & \cite{marzianietal23} \\
Fairall 9 & B1 & 0.04609 & 4550 & 0.43 & $2 \times 10^{45}$ & 1.5 $\cdot 10^8$ & 0.01 & 0.92 & $R_\mathrm{K}$: \cite{sikoraetal07}, \mbh: \cite{jiangetal21} \\
Mrk 335 & A1 & 0.0256 & 2175 & 0.34 & $6.9 \times 10^{44}$ & $3.45 \times 10^{7}$ & 0.13 & 0.36 \\ 
Mrk 478 & A2/A3 & 0.077 & 1322 & 1.04 & $4.4 \times 10^{45}$ & $2.90 \times 10^{7}$ & 1.02 & 0.85 & \cite{marshalletal96,hwangetal97,yuanetal04} \\
%Ark 120 & B1 & 0.033 & 5863 & $1.71 \times 10^{45}$ & $3.52 \times 10^{8}$ & 0.03 & 0.18 \\ 
PHL 1092 & A4 & 0.3965 & 2494 & 1.76 & $1.4 \times 10^{46}$ & $2.21 \times 10^{8}$ & 0.42 & 0.78 & Radio-detected but RQ \cite{zachariasetal05,marinelloetal20b} \\
\hline
Composite & xA & 2.1 - 2.5 & 3200$^b$ & \ldots$^b$ & $4\times 10^{46} - 4 \times 10^{47}$ & $ 10^9 - 10^{10}$\ & 0.7 -- 3 & $\lesssim$ 80 & $<$\lledd$>$ $\approx 1; $\cite{sniegowskaetal21,garnicaetal22}\\
\hline 
\end{tabular}
\caption*{\footnotesize {\bf Note.} (1) Object common name. (2) Spectral type of the object. (3) Redshift. (4) Full Width at Half Maximum of H$\beta$ expressed in units of km s$^{-1}$. (5) \feii\ prominence parameter, defined as the intensity ratio \rfe = \feiiq/\hb. (6) Bolometric luminosity ($L_{\rm bol}$) obtained by integrating the specific flux obtained from the NED database over all frequencies. (7) Mass of the supermassive black hole (M$_{\rm BH}$) of the AGN, calculated using the scaling relation from \cite{vestergaard06}. (8) Ratio between the bolometric luminosity of the object and the Eddington luminosity: \lledd = $L_{\rm bol}/[{1.5 \times 10^{38} {(M_{\rm BH}}/{M_\odot}})] $. (9) Kellerman Ratio obtained from $\frac{{\rm f}_{\text{Radio}}}{{\rm f}_{B}}$\ \cite{kellermannetal89}, where f$_{\text{Radio}}$ is the specific flux at a wavelength of 6 cm (5 GHz) and f$_{B}$ is the specific flux at 4400 \AA\ (680 THz) in the B band. $^a$: not detected, only a broad upper limit is estimated. $^b$: \hb\ spectral range not covered: \rfe\ not available. The FWHM refers to the best proximate of \hb\ in the UV, the \aliii\ doublet. 
}
\end{table}

\section{Results}
\label{results}
\subsection{Line profile analysis}

The results of the line profile analysis are shown, as an example, for the case of Fairall 9 (Fig. \ref{fig:f9fits}). All the lines fit with the three components introduced in Section \ref{multi}. \heiiuv\ and \civ\ are explained assuming that the components are present with consistent shifts and widths but changing their relative intensity ratios. This accounts for the flat \heiiuv\ profile that lacks the prominent BC observed for \civ, in turn implying \civ/\heiiuv\ $\gg$1\ for the BC. The VBC \civ/\heiiuv\ is much lower. A similar effect is visible for \hb\ and \heiiopt\ in the rightmost panel of Fig. \ref{fig:f9fits}. This has implications on the physical conditions derived for the VBLR and BLR, in turn motivating the model with two separate components, as discussed in several works \cite[][]{popovicetal04,sneddengaskell07,bonetal09}. A second important implication is that the VBC of \heiiopt\ and \heiiuv\ is much better defined than the BC, allowing for a more reliable estimate of the VBC ratios involving these lines. 

Similar line profile decomposition has been carried out for quasar emission lines over a broad range of redshift \cite{leighlymoore04,vietrietal17,yangetal23}, although the heuristic technique applied in this and some previous papers allows to consistently fit all quasars emission lines with only three components. The approach is equivalent to measuring profile intensity ratios \cite{shangetal07,garnicaetal22}, with the advantage that the absorptions that are frequently found in high-redshift quasar spectra can be easily compensated. Intensity ratios for the three components are found consistent with the ones of previous work. 

\subsection{Estimation of Metallicity for the Virialized Emitting Region}

\begin{figure}[t!]
\includegraphics[width=7.1 cm]{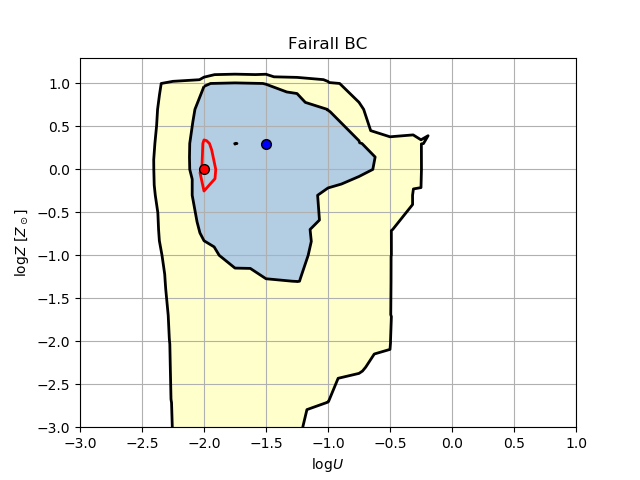}\includegraphics[width=7.1 cm]{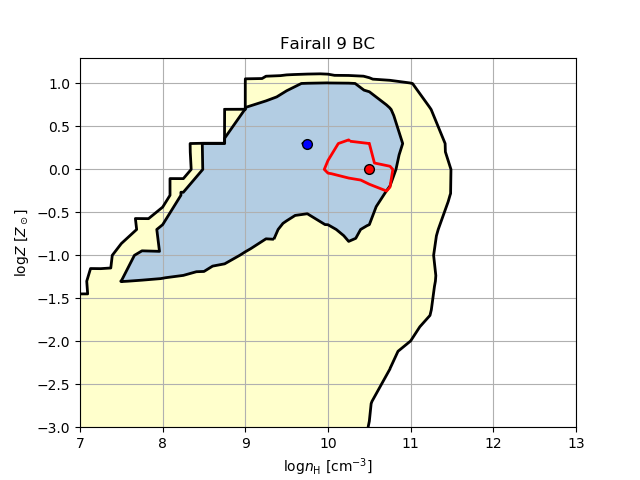}
\includegraphics[width=7.1 cm]{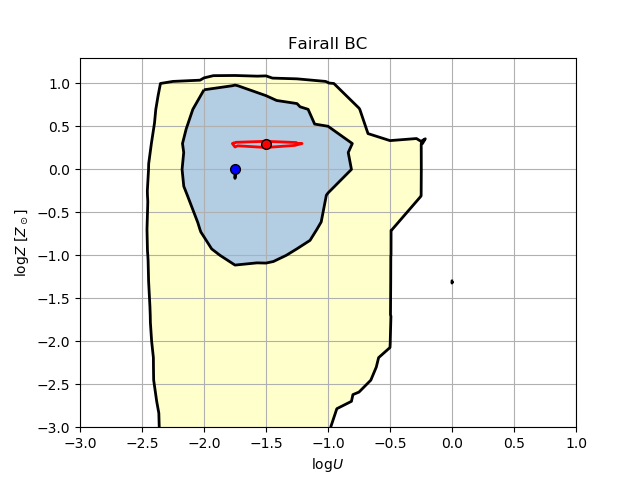}\includegraphics[width=7.1 cm]{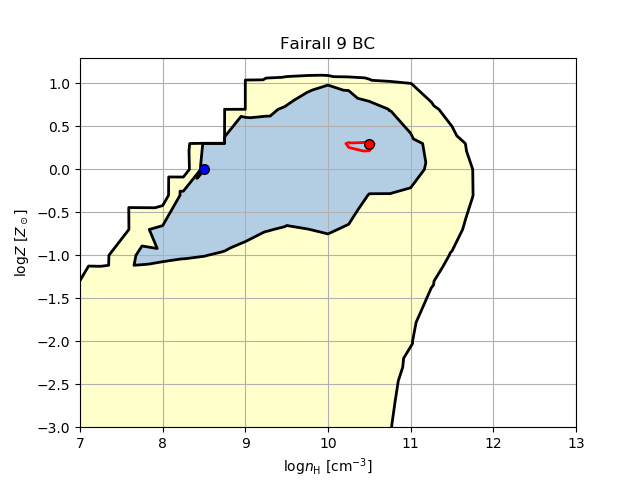}
\caption{\footnotesize Projections of the 3D parameter space ($U$, \nh, $Z$) onto the ($U$, $Z$) (left) and (\nh, $Z$) planes (right), for the prototypical Pop. B source Fairall 9. The top and bottom panels are for $\chi^2$\ computed with and without the \civ/\hb\ ratio. The blue spot identifies the model yielding minimum $\chi^2$\ for the BC; the red one is the same for the VBC. Isophotal contours are a 1$\sigma$ \ (pale blue) and 90\%\ confidence level (yellow); the red line is the 1$\sigma $\ level for the VBC. \label{fig:f9}}
\end{figure}

\begin{figure}[t!]
 \includegraphics[width=4.9 cm]{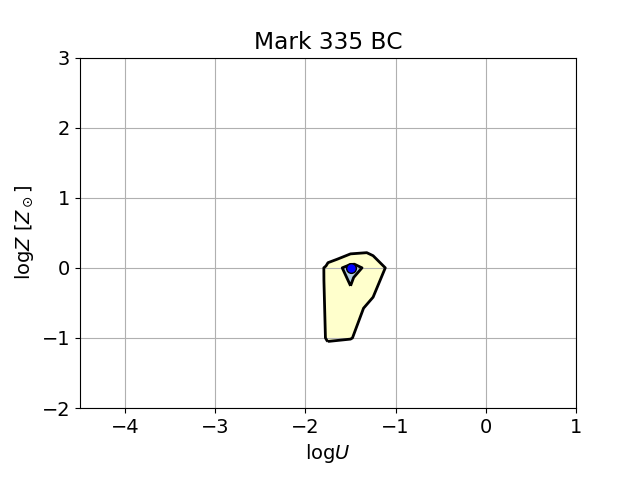}\hspace{-0.55cm}
 \includegraphics[width=4.9 cm]{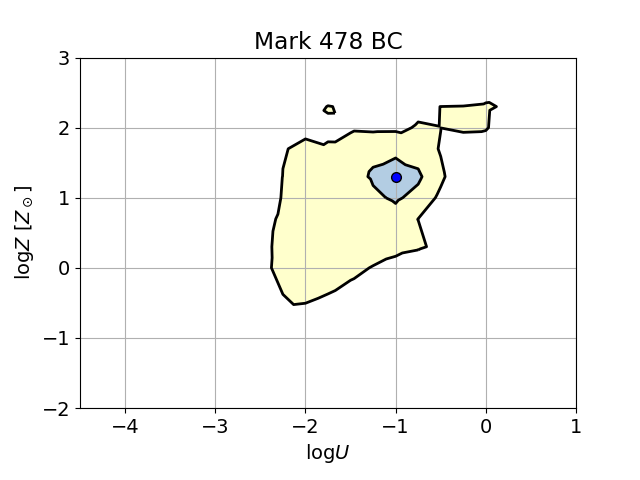}\hspace{-0.55cm}
 \includegraphics[width=4.9 cm]{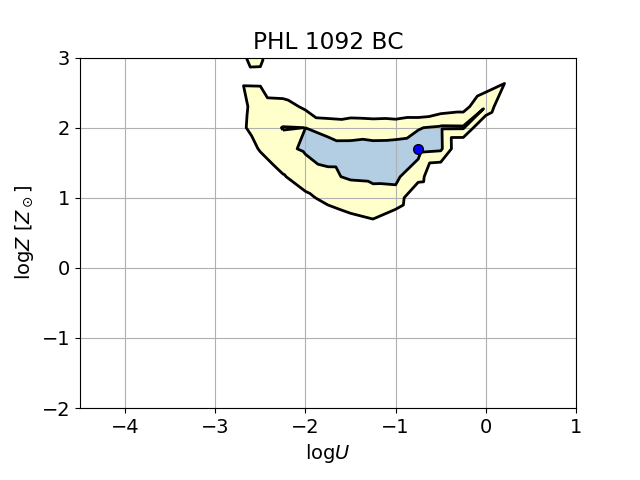}\\
 \includegraphics[width=4.9 cm]{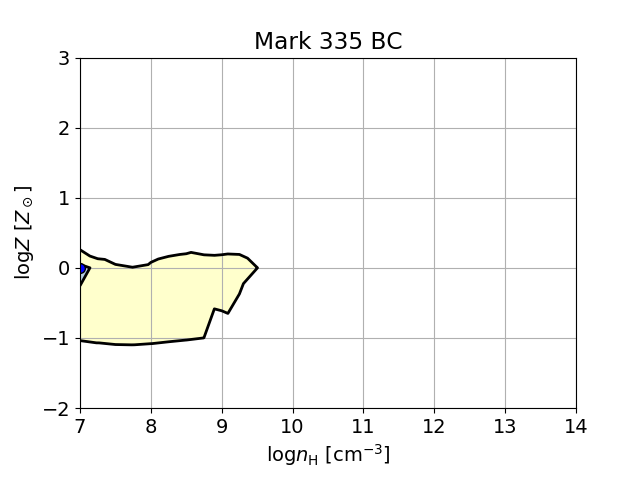}\hspace{-0.55cm}
\includegraphics[width= 4.9 cm]{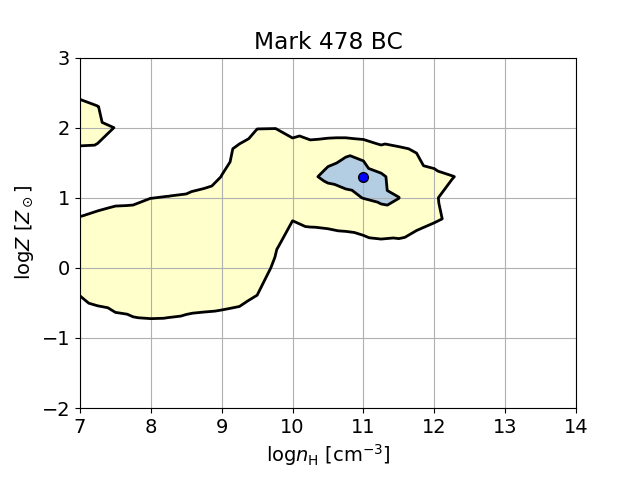}\hspace{-0.55cm} 
\includegraphics[width= 4.9 cm]{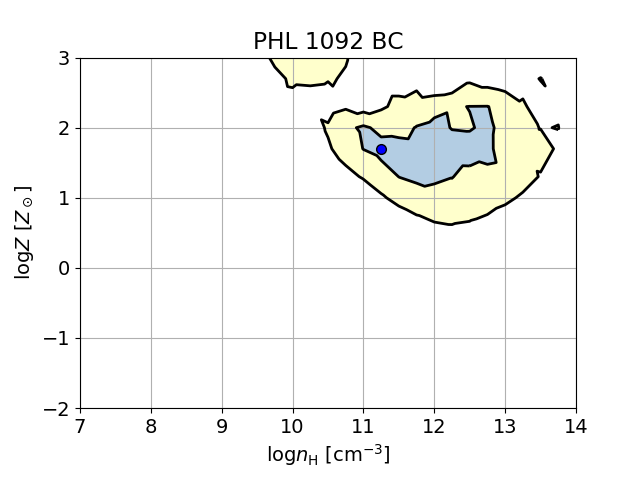} \\
\caption{\footnotesize Projections of the 3D parameter space ($U$, \nh, $Z$) onto the ($U$, $Z$) (top) and (\nh, $Z$) planes (bottom), for the Pop. A sources - Mark 335 (left), Mark 478 (middle), and PHL 1092 (right). The meaning of panels is the same as for the BC in Fig. \ref{fig:f9}. %All panels are for $\chi^2$\ computed without the \civ/\hb\ ratio, with 5 d.o.f. 
\label{fig:popa}}
\end{figure}

Four of the case studies are new results on individual objects, and the metallicity values and the associated uncertainties at $1 \sigma$\ confidence level are reported in Table \ref{tab:z}, along with estimates from previously published studies. Figs. \ref{fig:f9} and \ref{fig:popa} show the interval of confidence at 1$\sigma$\ level for the new cases in the planes $Z$ vs. ionization parameter and density \nh. 

The case of Fairall 9 includes the best $Z$\ derived for the VLBR. The $\chi^2$\ is lower even if the number of degrees of freedom is lower, which allows for a much more restricted range in the parameter space than for the case of the BLR ratios. In this case, the parameters are rather loosely constrained, although the agreement between the minimum $\chi^2$ \ derived for the BC and the one of the VBC, reinforces a metallicity estimate around $Z \approx 1 - 2 Z_\odot$, as the two regions are expected to have the same $Z$. In the Fairall 9 case, the consideration of the ratio \civ/\hb\ (an important parameter connected, in addition to $Z$, \ also to $U$), confirms the estimate $Z \approx 1 - 2 Z_\odot$, for both the BLR and VBLR. 

Fig. \ref{fig:popa} shows the planes $Z$ vs. ionization parameter and density \nh\ ordered along the sequence of increasing \rfe\ within Pop. A. The $Z$ values range from $\lesssim$ 0.1 Z$_{\odot}$ to 50 Z$_{\odot}$ from the { B1 RL composite} to PHL 1092. The $Z$\ value obtained for PHL 1092 confirms the high $Z \sim$ 100 Z$_{\odot}$ obtained with UV intensity ratios only \cite{sniegowskaetal21,garnicaetal22}. 

\unskip

\begin{figure}[t!]
\centering
\includegraphics[width=12. cm]{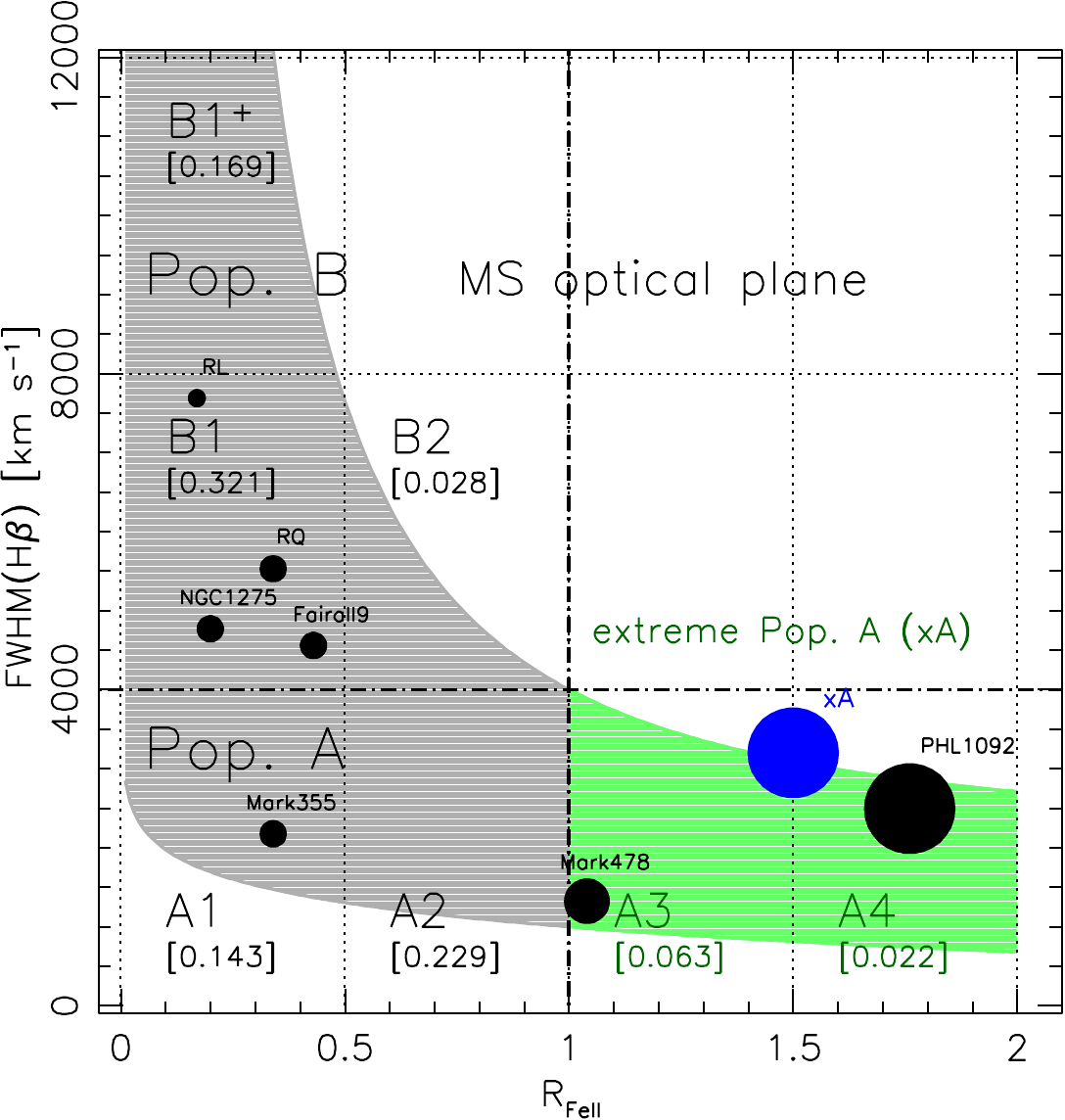}\vspace{-0.5cm}\\
\caption{\footnotesize Sketch of the quasar main sequence at low redshift, with circles of different sizes representing the metallicity estimates for different sources, composites, and samples presented in this paper and analyzed in recent literature. The abscissa is \feii\ prominence parameter \rfe; ordinate is the FWHM \hb\ in \kms. The numbers in square brackets report the prevalence of each spectral bin in an SDSS - based sample \cite{marzianietal13a}. The size of the circles depends on $Z$: smallest for $Z \lesssim 0.1 Z_\odot$, intermediate for $0.5 Z_\odot \lesssim Z \lesssim 2 Z_\odot$, larger for $Z \sim 10 Z_\odot$, and largest for $Z \sim 50 Z_\odot$. The blue circle refers to the intermediate $z$\ sample of \citet{garnicaetal22}. \label{fig:e1f}}
\end{figure}

Fig. \ref{fig:e1f} shows the location of all case studies along the E1 main sequence. The $Z$ values along the main sequence range from $0.1 Z_\odot$\ to $\sim$ 100 $Z_\odot$, and the trend is one of a monotonic increase along the horizontal sequence of increasing \rfe, from solar or slightly subsolar, to highly supersolar, with $Z$\ at least a few tens the solar values. Spectral type B1 and A1 consistently show similar values around solar, with the weak but still significant \feii\ emission. 

An important result is the realization that not all BLRs are made of gas with the same metal content. There is apparently a systematic gradient involving a range of more than a factor $\sim 100$. No matter the exact values of the $Z$, especially at the extremes, the trend is substantiated by the trend in the most metal-sensitive ratios, (\siiv+\oiv)/\civ, and \rfe, and ratios involving \aliii. The range in $Z$ can be compared with a recent systematic study for intermediate redshift quasars \cite{xuetal18}. Highest values around $Z \sim 20 Z_\odot$\ are found for massive black holes ($\log $\mbh\ $\sim 9.7$\ [\msol]), and are comparable with the value we obtain at low-$z$\ for high Eddington ratio sources. 

At intermediate redshift, however, the BLR $Z$ remains always highly supersolar, $\gtrsim 5 Z_\odot$, much above the most metal-rich galaxies ($Z \sim 2 Z_\odot$) \cite{maiolinomannucci19}. Quasars in the local Universe with modest masses ($\lesssim 10^8$ [\msol]) radiating at low \lledd\ are not yet sampled in major surveys, and there is therefore no disagreement if low-$Z$\ sources are missing at intermediate redshift.

\unskip

\begin{table}[h!]
\footnotesize
\caption{Metallicity estimates along the quasar MS}
\label{tab:z} % is used to refer this table in the text
\centering % used for centering table
\begin{tabular}{l c c c l} % centered columns (4 columns)
\hline\hline % inserts double horizontal lines
Identification & Comp. & $\log [Z/Z_\odot]$ & $1\sigma$ range & Notes\\ 
(1) & (2) & (3) & (4) \\
\hline 
Composite RL B1 & BC & --1.7 & --2 --- --1 & \cite{marzianietal23} \\
NGC 1275 & BC & --0.3 & --0.3 --- 0 & \cite{punslyetal18} \\
Composite RQ B1 & BC & 0.30 & --0.70 --- 1.00 & \cite{marzianietal23} \\
Composite RQ B1 & VBC & 0.70 & 0.70 --- 0.70 & \cite{marzianietal23} \\
Fairall 9 & BC & 0.30 & --1.30 --- 1.00 & This work; \civ/\hb\ excluded \\
Fairall 9 & VBC & 0.00 & 0.00 --- 0.30 & This work; \civ/\hb\ excluded \\
Fairall 9 & BC & 0.00 & --1.00 --- 0.70 & This work \\
Fairall 9 & VBC & 0.30 & 0.30 --- 0.30 & This work \\
Mrk 335 & BC & 0.00 & --1.00 --- 0.00$^a$ & This work \\ 
Mrk 478 & BC & 1.30 & 1.00 --- 1.30 & This work \\ 
PHL 1092 & BC & 1.70 & 1.30 --- 2.30 & This work \\
\hline
Composite $z\sim 2$ & xA & 1.70 & 1.30 --- 2.00 & \cite{sniegowskaetal21,garnicaetal22}\\
\hline 
\end{tabular}
%\caption*{\footnotesize {\bf Note.} \swayam{\textbf{missing notes}} }
\end{table}

\subsection{Outflows Traced by the blueshifted components}
\label{blue}

Gas outflows appear to be a phenomenon shared by the vast majority of AGNs \cite[e.g.,][]{feruglioetal10,harrisonetal14,feruglioetal15,wooetal16,kovacevicdojcinovicetal22}. Emission lines like \civ$\lambda$1549 provide valuable information about massive, ionized outflows associated with the accretion disk, and in turn, contribute to understanding the dynamic processes within the BLR and its interplay with the accretion disk. 

Selection of the lines most suitable for plasma diagnostics, including the metallicity, is much harder for the outflow component than for the virialized one, as the first appears spectroscopically resolved in high- and low-ionization lines only for extreme Population A or at very high luminosity. We used four diagnostic ratios for estimating the physical properties of the blue, outflowing component: \civ/\heiiuv, \civ/(\oiv\ + \siiv), \civ/\hb\ and (\oiv +\siiv)/\heiiuv, along with constraints from upper limits for the ratios \aliii/\civ, \feii/\hb. The diagnostic ratios are consistent with the estimates of metallicity derived for the virialized components, for PHL 1092 ($Z \sim 50 Z_\odot$, with a 1$\sigma$\ confidence range 20 -- 50 $Z_\odot$). Median values of the BLUE components are $\sim 6 Z_\odot$ for the xA intermediate $z$\ samples, somewhat lower and obtained only with three diagnostic ratios (\hb\ was not measured). Note that this abundance estimate is obtained using ratios involving only $\alpha$-elements. Other properties of the outflowing component are less well-constrained because of the limited number of diagnostics. A better precision might be achieved with higher S/N and additional diagnostic ratios. However, it is reasonable to assume that the outflowing gas from the high Eddington ratio sources might have $Z \sim 10 Z_\odot$, a conservative estimate in agreement with other works \cite[][and references therein]{xuetal18}. If this is the case, the mass outflow rate is $\dot{M} \sim 10^3 L_\mathrm{CIV, 45}, n_\mathrm{H,9}^{-1}$, where $L_\mathrm{CIV, 45}$\ \ergss\ is the \civ\ luminosity in units of $10^{45}$\ \ergss\ and $n_\mathrm{H,9}$ is the Hydrogen density in units of 10$^9$ \cm3\ \cite{marzianietal17c}. Solar metallicity implies that 1.46\%\ of the mass of the gas is due to metals. For a modest luminosity of $10^{44}$\ \ergss, the implication is that about $\sim 15$ \msol/yr of {\em metals} are returned to the interstellar medium (ISM). Over a high-accretion lifetime of $\sim 10^7$ yr { \cite{collinetal02,li12}}, the metal mass expected to be returned to the ISM could be $\sim 10^8$ \msol. While this estimate is extremely coarse, and the actual effect will depend on how the outflow is dissipated within the host, it implies that there could be in principle a significant enrichment for a large stellar population and earlier for the ISM. 

We have focused on low-luminosity quasars in the local Universe, characterized by relatively small black holes radiating near their Eddington limit. However, it's important to note that the outflow phenomena are most prominent in the highest luminosity quasars. These super-luminous quasars exhibit a high prevalence of significant blueshifts in the blueshifts in the \civfull\ and \oiiiopt\ emission line profiles \cite{bischettietal17,vietrietal18,deconto-machadoetal23}. The ionized gas mass, kinetic energy, and mechanical thrust in these cases are remarkably high, implying extensive feedback effects on the host galaxies of these exceptionally luminous quasars. These effects were particularly pronounced during cosmic epochs between 2 and 6 billion years after the Big Bang, suggesting that these quasars might have played a substantial role in enriching the chemical composition of their host galaxies.

\section{Discussion}
\label{disc}
\subsection{A gradient in metal content and chemical feedback along the sequence}

\begin{figure}[t!]
\includegraphics[width=14 cm]{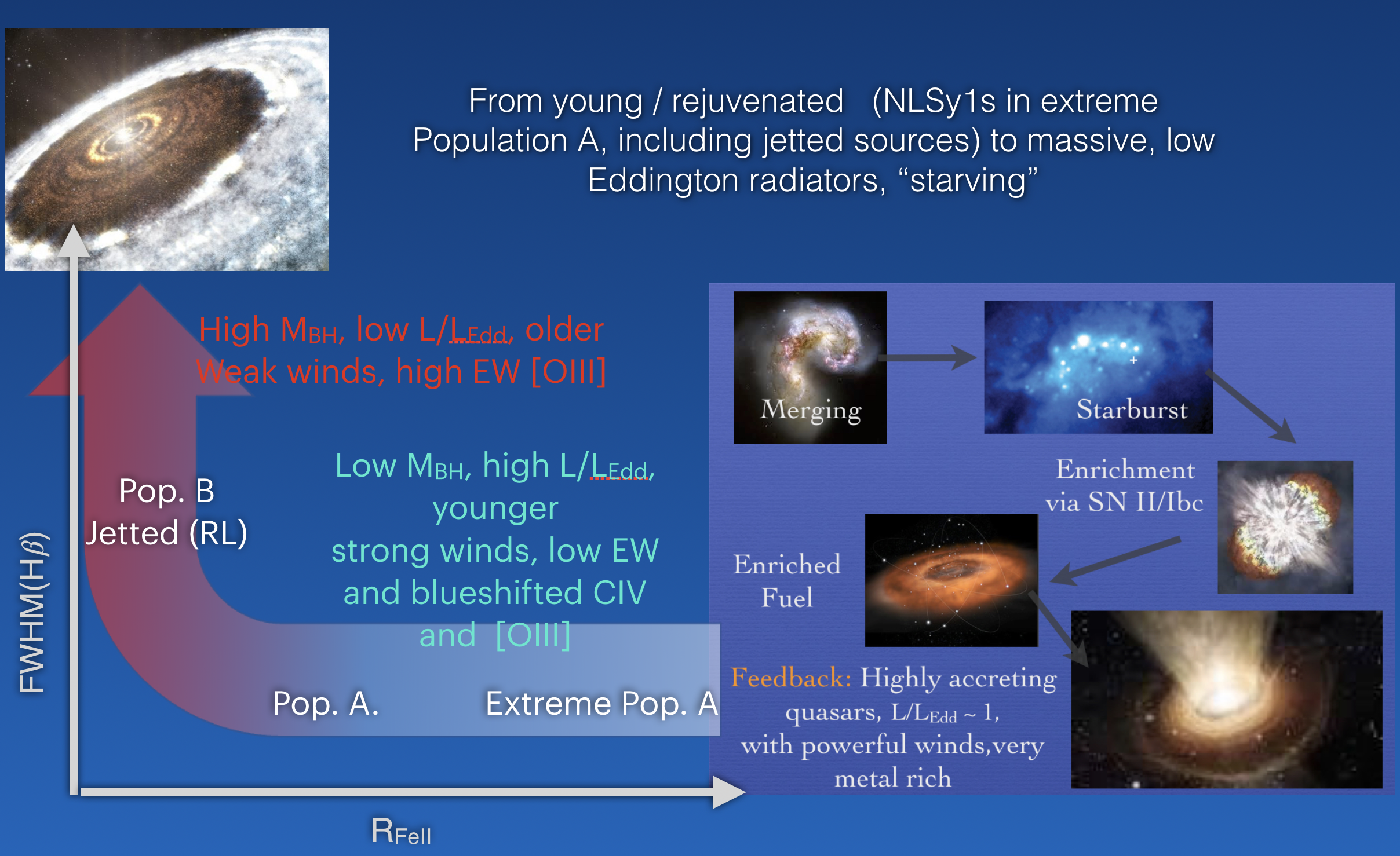}
\caption{Sketch depicting the evolutionary interpretation of the quasar main sequence. The inset on the right shows a possible evolutionary path leading to highly accreting quasar showing evidence of high metal enrichment in the BLR. \label{fig:evol}}
\end{figure}

 A cartoon depicting a global evolutionary scenario is shown in Fig. \ref{fig:evol}. At the one end of the sequence we encounter low mass, high Eddington ratio sources. They are accreting at a high rate, possibly following a merger and a burst of star formation. In the initial phases of the development of AGN and quasars, a series of events take place: wet mergers and strong interactions cause the accumulation of gas in the central regions of the galaxy. This accumulation triggers a burst of star formation (top of the inset) \cite{sandersetal88,sandersetal09}. Over time, mass loss from stellar winds and supernova explosions provide a source of accretion fuel for the massive black hole located at the center of the galaxy. Subsequently, radiation force and mechanical energy can clear away the dust surrounding the black hole, particularly within a cone aligned with the axis of the accretion disk. This allows the radiative and mechanical energy to be released into the ISM of the host galaxy (bottom of the inset). 
 %\chony{It is difficult to follow the thread of reasoning, and even more so with the mention to the top and bottom inserts}. 
 
 At a high accretion rate, a nuclear starburst can occur in the region where the disk becomes self-gravitating \citep{collinzahn99}. As mentioned, the $Z$-values calculated for the BLR of quasars appear exceptionally high when compared to their host galaxies. For reference, the highest measured $Z$-value in a molecular cloud is approximately 5 times the solar metallicity \cite{maiolinomannucci19}. However, it's important to note that the nuclear and circumnuclear environments of quasars may exhibit significant deviations from a typical interstellar setting. In these regions, stars traverse the disk, giving rise to the formation of accretion-modified objects that eventually attain substantial mass and, after a brief evolutionary phase, explode as core-collapse supernovae \citep{collinzahn99,chengwang99}. Stars within the nuclear vicinity can rapidly become highly massive (with masses exceeding 100 times that of the Sun), leading to core-collapse events that contribute to enriching the disk with heavy elements via the substantial metal yields produced by supernova ejecta \cite{cantielloetal21}. Furthermore, the compact remnants of these stars may continue accreting material, resulting in recurrent supernova occurrences \cite{lin97}. These accretion-modified star formation processes that enhance metallicity are projected to yield metal abundances approximately in the range of 10 to 20 times solar metallicity \cite{wangetal11}, which aligns with the values observed in the xA sources.

At the other end, we find very massive black holes, radiating at low Eddington ratios, in conditions that are proximate to ``starvation" and in any case to the exhaustion of the reservoir of gas for accretion. Accretion material may come from evolved star winds that can sustain modeled accretion rates \citep{padovanimatteucci93}. Metallicity values might be in this case more conventional, around solar or subsolar, such as the values { ranging from $\sim$0.1 to $\approx$ 2 times the solar metallicity found in the bulge of the Milky Way } \cite{zoccalietal03,gonzalezgadotti16}.

{ Pop. B sources have a slightly larger redshift than Pop. A ($\overline{z} \approx 0.33$ vs.  0.07) in a large, SDSS-based, low-$z$ sample \cite{zamfiretal10} that might be easily explained by selection effects. However, a ten-fold increase in black hole mass from 10$^{8}$ M$_{\odot}$\ at moderate accretion rate can occur over a time $ \sim 5 \cdot 10^{8}$ yr \cite{fraix-burnetetal17}. This timescale would correspond to a change in redshift $\delta z \approx 0.05$. Therefore, it is reasonable to assume that at least some Pop. B AGN evolved from Pop. A sources locally. } 

\unskip

\section{Conclusion}

In conclusion, the main sequence concept in quasars reveals a gradient of metallicity across different quasar populations, shedding light on their evolutionary paths. The understanding of these populations and their associated characteristics contributes to a deeper comprehension of the processes driving quasar behavior and their influence on their host galaxies. Expectations from accretion-modified stars within the active nuclei are consistent with highly supersolar metallicity.

%The complexities revealed through studies in this field continue to shape our understanding of the universe's most energetic phenomena.

%%%%%%%%%%%%%%%%%%%%%%%%%%%%%%%%%%%%%%%%%%

%%%%%%%%%%%%%%%%%%%%%%%%%%%%%%%%%%%%%%%%%%
\vspace{6pt} 

%%%%%%%%%%%%%%%%%%%%%%%%%%%%%%%%%%%%%%%%%%
%% optional
%\supplementary{The following supporting information can be downloaded at: \linksupplementary{s1}, Figure S1: title; Table S1: title; Video S1: title.}


%\begin{tabularx}{\textwidth}{lll}
%\toprule
%\textbf{Name} & \textbf{Type} & \textbf{Description} \\
%\midrule
%S1 & Python script (.py) & Script of python source code used in XX \\
%S2 & Text (.txt) & Script of modelling code used to make Figure X \\
%S3 & Text (.txt) & Raw data from experiment X \\
%S4 & Video (.mp4) & Video demonstrating the hardware in use \\
%\bottomrule
%\end{tabularx}
%}

%%%%%%%%%%%%%%%%%%%%%%%%%%%%%%%%%%%%%%%%%%
\authorcontributions{
Conceptualization, P. Marziani, S. Panda; methodology: P. Marziani, M. \'Sniegowska, S. Panda; formal analysis, P. Marziani, S. Panda, M. \'Sniegowska; Resources, D. Dultzin, A. del Olmo, M. D'Onofrio; data curation/analysis, A. Floris, K. Garnica, M. \'Sniegowska, A. Deconto-Machado; original draft preparation: P. Marziani; writing--review and editing: A. Floris, A. Deconto-Machado, M. \'Sniegowska, S. Panda, M. D'Onofrio, E. Bon, N. Bon, A. del Olmo, M. D'Onofrio; funding acquisition, A. del Olmo, D. Dultzin. All authors have read and agreed to the published version of the manuscript. 

}
%For research articles with several authors, a short paragraph specifying their individual contributions must be provided. The following statements should be used ``Conceptualization, X.X. and Y.Y.; methodology, X.X.; software, X.X.; validation, X.X., Y.Y. and Z.Z.; formal analysis, X.X.; investigation, X.X.; resources, X.X.; data curation, X.X.; writing---original draft preparation, X.X.; writing---review and editing, X.X.; visualization, X.X.; supervision, X.X.; project administration, X.X.; funding acquisition, Y.Y. All authors have read and agreed to the published version of the manuscript.'', please turn to the \href{http://img.mdpi.org/data/contributor-role-instruction.pdf}{CRediT taxonomy} for the term explanation. Authorship must be limited to those who have contributed substantially to the work~reported.}

\funding{This research received no external funding.}

\institutionalreview{Not applicable.}

\informedconsent{Not applicable.}

\dataavailability{Data available on request from the corresponding author.} 

% Only for journal Nursing Reports
%\publicinvolvement{Please describe how the public (patients, consumers, carers) were involved in the research. Consider reporting against the GRIPP2 (Guidance for Reporting Involvement of Patients and the Public) checklist. If the public were not involved in any aspect of the research add: ``No public involvement in any aspect of this research''.}

% Only for journal Nursing Reports
%\guidelinesstandards{Please add a statement indicating which reporting guideline was used when drafting the report. For example, ``This manuscript was drafted against the XXX (the full name of reporting guidelines and citation) for XXX (type of research) research''. A complete list of reporting guidelines can be accessed via the equator network: \url{https://www.equator-network.org/}.}

\acknowledgments{SP acknowledges the Conselho Nacional de Desenvolvimento Científico e Tecnológico (CNPq) Fellowships (164753/2020-6 and 313497/2022-2).
NB and EB acknowledge the support of the Serbian Ministry of Education, Science, and Technological Development, through the 695 contract number 451-03-68/2024-14/20000. AdO and ADM acknowledge financial support from the Spanish MCIN through project PID2022-140871NB-C21 and the Severo Ochoa grant CEX2021-001131-S funded by MCIN/AEI/10.13039/501100011033.}

\conflictsofinterest{The authors declare no conflict of interest.} 

%%%%%%%%%%%%%%%%%%%%%%%%%%%%%%%%%%%%%%%%%%
%% Optional
%\sampleavailability{Samples of the compounds ... are available from the authors.}

%% Only for journal Encyclopedia
%\entrylink{The Link to this entry published on the encyclopedia platform.}

\abbreviations{Abbreviations}{
The following abbreviations are used in this manuscript:\\

\noindent 
\begin{tabular}{@{}ll}
4DE1 & Four Dimensional Eigenvector 1\\
AGN & Active Galactic Nuclei\\
BLR & Broad Line Region\\
COS & Cosmic Origin Spectrograph\\
FIR & Far Infrared\\
FOS & Faint Object Spectrograph\\
FWHM & Full Width at Half Maximum\\
HIL & High Ionization Line\\
HST & Hubble Space Telescope\\
LIL & Low Ionization Line\\
MAST & Mikulski Archive of Space Telescopes\\
MS & Main Sequence\\
NGC & New General Catalog\\
SED & Spectral Energy Distribution\\
ST & Spectral Type\\
RL & Radio Loud\\
RQ & Radio Quiet\\
SOAR & Southern Observatory for Astrophysical Research \\
UV & Ultraviolet
\end{tabular}
}

%%%%%%%%%%%%%%%%%%%%%%%%%%%%%%%%%%%%%%%%%%
%% Optional
%\appendixtitles{no} % Leave argument "no" if all appendix headings stay EMPTY (then no dot is printed after "Appendix A"). If the appendix sections contain a heading then change the argument to "yes".
%\appendixstart
%\appendix
%\section[\appendixname~\thesection]{}
%\subsection[\appendixname~\thesubsection]{}

%\section[\appendixname~\thesection]{}

%%%%%%%%%%%%%%%%%%%%%%%%%%%%%%%%%%%%%%%%%%
\begin{adjustwidth}{-\extralength}{0cm}

\reftitle{References}
\externalbibliography{yes}
%\begin{thebibliography}{-------}

\bibliography{biblioletter3}

\begin{thebibliography}{999}

\bibitem[{Boroson} and {Green}(1992)]{borosongreen92}
{Boroson}, T.A.; {Green}, R.F.
\newblock {The Emission-Line Properties of Low-Redshift Quasi-stellar Objects}.
\newblock {\em \apjs} {\bf 1992}, {\em 80},~109.
\newblock {\url{https://doi.org/10.1086/191661}}.

\bibitem[{Gaskell}(1985)]{gaskell85}
{Gaskell}, C.M.
\newblock {Galactic mergers, starburst galaxies, quasar activity and massive
  binary black holes}.
\newblock {\em \nat} {\bf 1985}, {\em 315},~386--+.
\newblock {\url{https://doi.org/10.1038/315386a0}}.

\bibitem[{Sulentic} \em{et~al.}(2000){Sulentic}, {Zwitter}, {Marziani}, and
  {Dultzin-Hacyan}]{sulenticetal00c}
{Sulentic}, J.W.; {Zwitter}, T.; {Marziani}, P.; {Dultzin-Hacyan}, D.
\newblock {Eigenvector 1: An Optimal Correlation Space for Active Galactic
  Nuclei}.
\newblock {\em ApJL} {\bf 2000}, {\em 536},~L5--L9,
  \href{http://xxx.lanl.gov/abs/arXiv:astro-ph/0005177}{{\normalfont
  [arXiv:astro-ph/0005177]}}.
\newblock {\url{https://doi.org/10.1086/312717}}.

\bibitem[{Zamfir} \em{et~al.}(2010){Zamfir}, {Sulentic}, {Marziani}, and
  {Dultzin}]{zamfiretal10}
{Zamfir}, S.; {Sulentic}, J.W.; {Marziani}, P.; {Dultzin}, D.
\newblock {Detailed characterization of H{$\beta$} emission line profile in
  low-z SDSS quasars}.
\newblock {\em \mnras} {\bf 2010}, {\em 403},~1759,
  \href{http://xxx.lanl.gov/abs/0912.4306}{{\normalfont [0912.4306]}}.
\newblock {\url{https://doi.org/10.1111/j.1365-2966.2009.16236.x}}.

\bibitem[{Shen} and {Ho}(2014)]{shenho14}
{Shen}, Y.; {Ho}, L.C.
\newblock {The diversity of quasars unified by accretion and orientation}.
\newblock {\em \nat} {\bf 2014}, {\em 513},~210--213,
  \href{http://xxx.lanl.gov/abs/1409.2887}{{\normalfont [1409.2887]}}.
\newblock {\url{https://doi.org/10.1038/nature13712}}.

\bibitem[{Sun} and {Shen}(2015)]{sunshen15}
{Sun}, J.; {Shen}, Y.
\newblock {Dissecting the Quasar Main Sequence: Insight from Host Galaxy
  Properties}.
\newblock {\em \apjl} {\bf 2015}, {\em 804},~L15,
  \href{http://xxx.lanl.gov/abs/1503.08364}{{\normalfont [1503.08364]}}.
\newblock {\url{https://doi.org/10.1088/2041-8205/804/1/L15}}.

\bibitem[{Panda} \em{et~al.}(2019){Panda}, {Marziani}, and
  {Czerny}]{pandaetal19}
{Panda}, S.; {Marziani}, P.; {Czerny}, B.
\newblock {The Quasar Main Sequence Explained by the Combination of Eddington
  Ratio, Metallicity, and Orientation}.
\newblock {\em \apj} {\bf 2019}, {\em 882},~79,
  \href{http://xxx.lanl.gov/abs/1905.01729}{{\normalfont
  [arXiv:astro-ph.HE/1905.01729]}}.
\newblock {\url{https://doi.org/10.3847/1538-4357/ab3292}}.

\bibitem[{Giustini} and {Proga}(2019)]{giustiniproga19}
{Giustini}, M.; {Proga}, D.
\newblock {A global view of the inner accretion and ejection flow around super
  massive black holes. Radiation-driven accretion disk winds in a physical
  context}.
\newblock {\em \aap} {\bf 2019}, {\em 630},~A94,
  \href{http://xxx.lanl.gov/abs/1904.07341}{{\normalfont
  [arXiv:astro-ph.GA/1904.07341]}}.
\newblock {\url{https://doi.org/10.1051/0004-6361/201833810}}.

\bibitem[{Du} \em{et~al.}(2016){Du}, {Lu}, {Hu}, {Qiu}, {Li}, {Huang}, {Wang},
  {Bai}, {Bian}, {Yuan}, {Ho}, {Wang}, and {SEAMBH Collaboration}]{duetal16}
{Du}, P.; {Lu}, K.X.; {Hu}, C.; {Qiu}, J.; {Li}, Y.R.; {Huang}, Y.K.; {Wang},
  F.; {Bai}, J.M.; {Bian}, W.H.; {Yuan}, Y.F.;  et~al.
\newblock {Supermassive Black Holes with High Accretion Rates in Active
  Galactic Nuclei. VI. Velocity-resolved Reverberation Mapping of the
  H{$\beta$} Line}.
\newblock {\em \apj} {\bf 2016}, {\em 820},~27,
  \href{http://xxx.lanl.gov/abs/1602.01922}{{\normalfont [1602.01922]}}.
\newblock {\url{https://doi.org/10.3847/0004-637X/820/1/27}}.

\bibitem[Fraix-Burnet \em{et~al.}(2017)Fraix-Burnet, Marziani, D'Onofrio, and
  Dultzin]{fraix-burnetetal17}
Fraix-Burnet, D.; Marziani, P.; D'Onofrio, M.; Dultzin, D.
\newblock The Phylogeny of Quasars and the Ontogeny of Their Central Black
  Holes.
\newblock {\em Frontiers in Astronomy and Space Sciences} {\bf 2017}, {\em
  4},~1.
\newblock {\url{https://doi.org/10.3389/fspas.2017.00001}}.

\bibitem[{Panda}(2021)]{Panda_2021}
{Panda}, S.
\newblock {The CaFe project: Optical Fe II and near-infrared Ca II triplet
  emission in active galaxies: simulated EWs and the co-dependence of cloud
  size and metal content}.
\newblock {\em \aap} {\bf 2021}, {\em 650},~A154,
  \href{http://xxx.lanl.gov/abs/2004.13113}{{\normalfont
  [arXiv:astro-ph.GA/2004.13113]}}.
\newblock {\url{https://doi.org/10.1051/0004-6361/202140393}}.

\bibitem[{Hamann} and {Ferland}(1993)]{hamannferland93}
{Hamann}, F.; {Ferland}, G.
\newblock {The Chemical Evolution of QSOs and the Implications for Cosmology
  and Galaxy Formation}.
\newblock {\em \apj} {\bf 1993}, {\em 418},~11.
\newblock {\url{https://doi.org/10.1086/173366}}.

\bibitem[{Nagao} \em{et~al.}(2006){Nagao}, {Marconi}, and
  {Maiolino}]{nagaoetal06}
{Nagao}, T.; {Marconi}, A.; {Maiolino}, R.
\newblock {The evolution of the broad-line region among SDSS quasars}.
\newblock {\em A\&Ap} {\bf 2006}, {\em 447},~157--172,
  \href{http://xxx.lanl.gov/abs/arXiv:astro-ph/0510385}{{\normalfont
  [arXiv:astro-ph/0510385]}}.
\newblock {\url{https://doi.org/10.1051/0004-6361:20054024}}.

\bibitem[{Juarez} \em{et~al.}(2009){Juarez}, {Maiolino}, {Mujica}, {Pedani},
  {Marinoni}, {Nagao}, {Marconi}, and {Oliva}]{juarezetal09}
{Juarez}, Y.; {Maiolino}, R.; {Mujica}, R.; {Pedani}, M.; {Marinoni}, S.;
  {Nagao}, T.; {Marconi}, A.; {Oliva}, E.
\newblock {The metallicity of the most distant quasars}.
\newblock {\em A\&Ap} {\bf 2009}, {\em 494},~L25--L28,
  \href{http://xxx.lanl.gov/abs/0901.0974}{{\normalfont [0901.0974]}}.
\newblock {\url{https://doi.org/10.1051/0004-6361:200811415}}.

\bibitem[{Matsuoka} \em{et~al.}(2011){Matsuoka}, {Nagao}, {Marconi},
  {Maiolino}, and {Taniguchi}]{matsuokaetal11}
{Matsuoka}, K.; {Nagao}, T.; {Marconi}, A.; {Maiolino}, R.; {Taniguchi}, Y.
\newblock {The mass-metallicity relation of SDSS quasars}.
\newblock {\em \aap} {\bf 2011}, {\em 527},~A100,
  \href{http://xxx.lanl.gov/abs/1011.5811}{{\normalfont
  [arXiv:astro-ph.CO/1011.5811]}}.
\newblock {\url{https://doi.org/10.1051/0004-6361/201015584}}.

\bibitem[{Shin} \em{et~al.}(2013){Shin}, {Woo}, {Nagao}, and {Kim}]{shinetal13}
{Shin}, J.; {Woo}, J.H.; {Nagao}, T.; {Kim}, S.C.
\newblock {The Chemical Properties of Low-redshift QSOs}.
\newblock {\em \apj} {\bf 2013}, {\em 763},~58,
  \href{http://xxx.lanl.gov/abs/1211.6749}{{\normalfont
  [arXiv:astro-ph.CO/1211.6749]}}.
\newblock {\url{https://doi.org/10.1088/0004-637X/763/1/58}}.

\bibitem[{Sameshima} \em{et~al.}(2017){Sameshima}, {Yoshii}, and
  {Kawara}]{sameshimaetal17}
{Sameshima}, H.; {Yoshii}, Y.; {Kawara}, K.
\newblock {Chemical Evolution of the Universe at 0.7 < z < 1.6 Derived from
  Abundance Diagnostics of the Broad-line Region of Quasars}.
\newblock {\em \apj} {\bf 2017}, {\em 834},~203,
  \href{http://xxx.lanl.gov/abs/1611.06027}{{\normalfont
  [arXiv:astro-ph.GA/1611.06027]}}.
\newblock {\url{https://doi.org/10.3847/1538-4357/834/2/203}}.

\bibitem[{Wang} \em{et~al.}(2022){Wang}, {Jiang}, {Shen}, {Ho}, {Vestergaard},
  {Ba{\~n}ados}, {Willott}, {Wu}, {Zou}, {Yang}, {Wang}, {Fan}, and
  {Wu}]{wangetal22}
{Wang}, S.; {Jiang}, L.; {Shen}, Y.; {Ho}, L.C.; {Vestergaard}, M.;
  {Ba{\~n}ados}, E.; {Willott}, C.J.; {Wu}, J.; {Zou}, S.; {Yang}, J.;  et~al.
\newblock {Metallicity in Quasar Broad-line Regions at Redshift 6}.
\newblock {\em \apj} {\bf 2022}, {\em 925},~121,
  \href{http://xxx.lanl.gov/abs/2112.07799}{{\normalfont
  [arXiv:astro-ph.GA/2112.07799]}}.
\newblock {\url{https://doi.org/10.3847/1538-4357/ac3a69}}.

\bibitem[{Matteucci}(2012)]{matteucci12}
{Matteucci}, F.
\newblock {\em {Chemical Evolution of Galaxies}}; Springer Verlag,  2012.
\newblock {\url{https://doi.org/10.1007/978-3-642-22491-1}}.

\bibitem[{Xu} \em{et~al.}(2018){Xu}, {Bian}, {Shen}, {Zuo}, {Fan}, and
  {Zhu}]{xuetal18}
{Xu}, F.; {Bian}, F.; {Shen}, Y.; {Zuo}, W.; {Fan}, X.; {Zhu}, Z.
\newblock {The evolution of chemical abundance in quasar broad line region}.
\newblock {\em \mnras} {\bf 2018}, {\em 480},~345--357,
  \href{http://xxx.lanl.gov/abs/1807.01978}{{\normalfont
  [arXiv:astro-ph.GA/1807.01978]}}.
\newblock {\url{https://doi.org/10.1093/mnras/sty1763}}.

\bibitem[{Marziani} \em{et~al.}(2023){Marziani}, {Panda}, {Deconto Machado},
  and {Del Olmo}]{marzianietal23}
{Marziani}, P.; {Panda}, S.; {Deconto Machado}, A.; {Del Olmo}, A.
\newblock {Metal Content in Relativistically Jetted and Radio-Quiet Quasars in
  the Main Sequence Context}.
\newblock {\em Galaxies} {\bf 2023}, {\em 11},~52,
  \href{http://xxx.lanl.gov/abs/2303.13250}{{\normalfont
  [arXiv:astro-ph.GA/2303.13250]}}.
\newblock {\url{https://doi.org/10.3390/galaxies11020052}}.

\bibitem[{{\'S}niegowska} \em{et~al.}(2021){{\'S}niegowska}, {Marziani},
  {Czerny}, {Panda}, {Mart{\'\i}nez-Aldama}, {del Olmo}, and
  {D'Onofrio}]{sniegowskaetal21}
{{\'S}niegowska}, M.; {Marziani}, P.; {Czerny}, B.; {Panda}, S.;
  {Mart{\'\i}nez-Aldama}, M.L.; {del Olmo}, A.; {D'Onofrio}, M.
\newblock {High Metal Content of Highly Accreting Quasars}.
\newblock {\em \apj} {\bf 2021}, {\em 910},~115,
  \href{http://xxx.lanl.gov/abs/2009.14177}{{\normalfont
  [arXiv:astro-ph.HE/2009.14177]}}.
\newblock {\url{https://doi.org/10.3847/1538-4357/abe1c8}}.

\bibitem[{Di Matteo} \em{et~al.}(2005){Di Matteo}, {Springel}, and
  {Hernquist}]{dimatteoetal05}
{Di Matteo}, T.; {Springel}, V.; {Hernquist}, L.
\newblock {Energy input from quasars regulates the growth and activity of black
  holes and their host galaxies}.
\newblock {\em \nat} {\bf 2005}, {\em 433},~604--607,
  \href{http://xxx.lanl.gov/abs/astro-ph/0502199}{{\normalfont
  [arXiv:astro-ph/astro-ph/0502199]}}.
\newblock {\url{https://doi.org/10.1038/nature03335}}.

\bibitem[{Hopkins} \em{et~al.}(2006){Hopkins}, {Hernquist}, {Cox}, {Di Matteo},
  {Robertson}, and {Springel}]{hopkinsetal06}
{Hopkins}, P.F.; {Hernquist}, L.; {Cox}, T.J.; {Di Matteo}, T.; {Robertson},
  B.; {Springel}, V.
\newblock {A Unified, Merger-driven Model of the Origin of Starbursts, Quasars,
  the Cosmic X-Ray Background, Supermassive Black Holes, and Galaxy Spheroids}.
\newblock {\em \apjs} {\bf 2006}, {\em 163},~1--49,
  \href{http://xxx.lanl.gov/abs/astro-ph/0506398}{{\normalfont
  [astro-ph/0506398]}}.
\newblock {\url{https://doi.org/10.1086/499298}}.

\bibitem[{Somerville} and {Dav{\'e}}(2015)]{somervilledave15}
{Somerville}, R.S.; {Dav{\'e}}, R.
\newblock {Physical Models of Galaxy Formation in a Cosmological Framework}.
\newblock {\em \araa} {\bf 2015}, {\em 53},~51--113,
  \href{http://xxx.lanl.gov/abs/1412.2712}{{\normalfont
  [arXiv:astro-ph.GA/1412.2712]}}.
\newblock {\url{https://doi.org/10.1146/annurev-astro-082812-140951}}.

\bibitem[{Marziani} \em{et~al.}(2003){Marziani}, {Sulentic}, {Zamanov},
  {Calvani}, {Dultzin-Hacyan}, {Bachev}, and {Zwitter}]{marzianietal03a}
{Marziani}, P.; {Sulentic}, J.W.; {Zamanov}, R.; {Calvani}, M.;
  {Dultzin-Hacyan}, D.; {Bachev}, R.; {Zwitter}, T.
\newblock {An Optical Spectroscopic Atlas of Low-Redshift Active Galactic
  Nuclei}.
\newblock {\em ApJS} {\bf 2003}, {\em 145},~199--211.
\newblock {\url{https://doi.org/10.1086/346025}}.

\bibitem[{Punsly} \em{et~al.}(2018){Punsly}, {Marziani}, {Bennert}, {Nagai},
  and {Gurwell}]{punslyetal18}
{Punsly}, B.; {Marziani}, P.; {Bennert}, V.N.; {Nagai}, H.; {Gurwell}, M.A.
\newblock {Revealing the Broad Line Region of NGC 1275: The Relationship to Jet
  Power}.
\newblock {\em \apj} {\bf 2018}, {\em 869},~143,
  \href{http://xxx.lanl.gov/abs/1810.11716}{{\normalfont
  [arXiv:astro-ph.GA/1810.11716]}}.
\newblock {\url{https://doi.org/10.3847/1538-4357/aaec75}}.

\bibitem[{Marinello} \em{et~al.}(2020){Marinello}, {Rodr{\'\i}guez-Ardila},
  {Marziani}, {Sigut}, and {Pradhan}]{marinelloetal20b}
{Marinello}, M.; {Rodr{\'\i}guez-Ardila}, A.; {Marziani}, P.; {Sigut}, A.;
  {Pradhan}, A.
\newblock {Panchromatic Properties of the Extreme Fe ii Emitter PHL 1092}.
\newblock {\em \mnras} {\bf 2020},
  \href{http://xxx.lanl.gov/abs/2004.01811}{{\normalfont
  [arXiv:astro-ph.GA/2004.01811]}}.
\newblock {\url{https://doi.org/10.1093/mnras/staa934}}.

\bibitem[{Garnica} \em{et~al.}(2022){Garnica}, {Negrete}, {Marziani},
  {Dultzin}, {{\'S}niegowska}, and {Panda}]{garnicaetal22}
{Garnica}, K.; {Negrete}, C.A.; {Marziani}, P.; {Dultzin}, D.;
  {{\'S}niegowska}, M.; {Panda}, S.
\newblock {High metal content of highly accreting quasars: Analysis of an
  extended sample}.
\newblock {\em \aap} {\bf 2022}, {\em 667},~A105,
  \href{http://xxx.lanl.gov/abs/2208.02387}{{\normalfont
  [arXiv:astro-ph.GA/2208.02387]}}.
\newblock {\url{https://doi.org/10.1051/0004-6361/202142837}}.

\bibitem[{Collin-Souffrin} \em{et~al.}(1988){Collin-Souffrin}, {Dyson},
  {McDowell}, and {Perry}]{collinsouffrinetal88}
{Collin-Souffrin}, S.; {Dyson}, J.E.; {McDowell}, J.C.; {Perry}, J.J.
\newblock {The environment of active galactic nuclei. I - A two-component broad
  emission line model}.
\newblock {\em MNRAS} {\bf 1988}, {\em 232},~539--550.

\bibitem[{Elvis}(2000)]{elvis00}
{Elvis}, M.
\newblock {A Structure for Quasars}.
\newblock {\em \apj} {\bf 2000}, {\em 545},~63--76,
  \href{http://xxx.lanl.gov/abs/arXiv:astro-ph/0008064}{{\normalfont
  [arXiv:astro-ph/0008064]}}.
\newblock {\url{https://doi.org/10.1086/317778}}.

\bibitem[{Peterson} and {Wandel}(1999)]{petersonwandel99}
{Peterson}, B.M.; {Wandel}, A.
\newblock {Keplerian Motion of Broad-Line Region Gas as Evidence for
  Supermassive Black Holes in Active Galactic Nuclei}.
\newblock {\em \apjl} {\bf 1999}, {\em 521},~L95--L98,
  \href{http://xxx.lanl.gov/abs/arXiv:astro-ph/9905382}{{\normalfont
  [arXiv:astro-ph/9905382]}}.
\newblock {\url{https://doi.org/10.1086/312190}}.

\bibitem[{Peterson} and {Wandel}(2000)]{petersonwandel00}
{Peterson}, B.M.; {Wandel}, A.
\newblock {Evidence for Supermassive Black Holes in Active Galactic Nuclei from
  Emission-Line Reverberation}.
\newblock {\em \apjl} {\bf 2000}, {\em 540},~L13--L16,
  \href{http://xxx.lanl.gov/abs/astro-ph/0007147}{{\normalfont
  [astro-ph/0007147]}}.
\newblock {\url{https://doi.org/10.1086/312862}}.

\bibitem[{Richards} \em{et~al.}(2011){Richards}, {Kruczek}, {Gallagher},
  {Hall}, {Hewett}, {Leighly}, {Deo}, {Kratzer}, and {Shen}]{richardsetal11}
{Richards}, G.T.; {Kruczek}, N.E.; {Gallagher}, S.C.; {Hall}, P.B.; {Hewett},
  P.C.; {Leighly}, K.M.; {Deo}, R.P.; {Kratzer}, R.M.; {Shen}, Y.
\newblock {Unification of Luminous Type 1 Quasars through C IV Emission}.
\newblock {\em \aj} {\bf 2011}, {\em 141},~167--+,
  \href{http://xxx.lanl.gov/abs/1011.2282}{{\normalfont
  [arXiv:astro-ph.GA/1011.2282]}}.
\newblock {\url{https://doi.org/10.1088/0004-6256/141/5/167}}.

\bibitem[{Coatman} \em{et~al.}(2016){Coatman}, {Hewett}, {Banerji}, and
  {Richards}]{coatmanetal16}
{Coatman}, L.; {Hewett}, P.C.; {Banerji}, M.; {Richards}, G.T.
\newblock {C iv emission-line properties and systematic trends in quasar black
  hole mass estimates}.
\newblock {\em \mnras} {\bf 2016}, {\em 461},~647--665,
  \href{http://xxx.lanl.gov/abs/1606.02726}{{\normalfont [1606.02726]}}.
\newblock {\url{https://doi.org/10.1093/mnras/stw1360}}.

\bibitem[{Ferland} \em{et~al.}(2009){Ferland}, {Hu}, {Wang}, {Baldwin},
  {Porter}, {van Hoof}, and {Williams}]{ferlandetal09}
{Ferland}, G.J.; {Hu}, C.; {Wang}, J.; {Baldwin}, J.A.; {Porter}, R.L.; {van
  Hoof}, P.A.M.; {Williams}, R.J.R.
\newblock {Implications of Infalling Fe II-Emitting Clouds in Active Galactic
  Nuclei: Anisotropic Properties}.
\newblock {\em \apjl} {\bf 2009}, {\em 707},~L82--L86,
  \href{http://xxx.lanl.gov/abs/0911.1173}{{\normalfont [0911.1173]}}.
\newblock {\url{https://doi.org/10.1088/0004-637X/707/1/L82}}.

\bibitem[{V{\'e}ron-Cetty} \em{et~al.}(2001){V{\'e}ron-Cetty}, {V{\'e}ron}, and
  {Gon{\c c}alves}]{veroncettyetal01}
{V{\'e}ron-Cetty}, M.P.; {V{\'e}ron}, P.; {Gon{\c c}alves}, A.C.
\newblock {A spectrophotometric atlas of Narrow-Line Seyfert 1 galaxies}.
\newblock {\em AAp} {\bf 2001}, {\em 372},~730--754,
  \href{http://xxx.lanl.gov/abs/arXiv:astro-ph/0104151}{{\normalfont
  [arXiv:astro-ph/0104151]}}.
\newblock {\url{https://doi.org/10.1051/0004-6361:20010489}}.

\bibitem[{Marziani} \em{et~al.}(2017){Marziani}, {Negrete}, {Dultzin},
  {Mart{\'{\i}}nez-Aldama}, {Del Olmo}, {D'Onofrio}, and
  {Stirpe}]{marzianietal17c}
{Marziani}, P.; {Negrete}, C.A.; {Dultzin}, D.; {Mart{\'{\i}}nez-Aldama}, M.L.;
  {Del Olmo}, A.; {D'Onofrio}, M.; {Stirpe}, G.M.
\newblock {Quasar massive ionized outflows traced by CIV {$\lambda$}1549 and
  [OIII]{$\lambda$}{$\lambda$}4959,5007}.
\newblock {\em Frontiers in Astronomy and Space Sciences} {\bf 2017}, {\em
  4},~16,  \href{http://xxx.lanl.gov/abs/1709.05691}{{\normalfont
  [1709.05691]}}.
\newblock {\url{https://doi.org/10.3389/fspas.2017.00016}}.

\bibitem[{Vietri} \em{et~al.}(2018){Vietri}, {Piconcelli}, {Bischetti},
  {Duras}, {Martocchia}, {Bongiorno}, {Marconi}, {Zappacosta}, {Bisogni},
  {Bruni}, {Brusa}, {Comastri}, {Cresci}, {Feruglio}, {Giallongo}, {La Franca},
  {Mainieri}, {Mannucci}, {Ricci}, {Sani}, {Testa}, {Tombesi}, {Vignali}, and
  {Fiore}]{vietrietal18}
{Vietri}, G.; {Piconcelli}, E.; {Bischetti}, M.; {Duras}, F.; {Martocchia}, S.;
  {Bongiorno}, A.; {Marconi}, A.; {Zappacosta}, L.; {Bisogni}, S.; {Bruni}, G.;
   et~al.
\newblock {The WISSH quasars project. IV. Broad line region versus
  kiloparsec-scale winds}.
\newblock {\em \aap} {\bf 2018}, {\em 617},~A81,
  \href{http://xxx.lanl.gov/abs/1802.03423}{{\normalfont [1802.03423]}}.
\newblock {\url{https://doi.org/10.1051/0004-6361/201732335}}.

\bibitem[{Deconto-Machado} \em{et~al.}(2023){Deconto-Machado}, {del Olmo
  Orozco}, {Marziani}, {Perea}, and {Stirpe}]{deconto-machadoetal23}
{Deconto-Machado}, A.; {del Olmo Orozco}, A.; {Marziani}, P.; {Perea}, J.;
  {Stirpe}, G.M.
\newblock {High-redshift quasars along the main sequence}.
\newblock {\em \aap} {\bf 2023}, {\em 669},~A83,
  \href{http://xxx.lanl.gov/abs/2211.03853}{{\normalfont
  [arXiv:astro-ph.GA/2211.03853]}}.
\newblock {\url{https://doi.org/10.1051/0004-6361/202243801}}.

\bibitem[{Leighly} and {Moore}(2004)]{leighlymoore04}
{Leighly}, K.M.; {Moore}, J.R.
\newblock {Hubble Space Telescope STIS Ultraviolet Spectral Evidence of Outflow
  in Extreme Narrow-Line Seyfert 1 Galaxies. I. Data and Analysis}.
\newblock {\em \apj} {\bf 2004}, {\em 611},~107--124,
  \href{http://xxx.lanl.gov/abs/arXiv:astro-ph/0402453}{{\normalfont
  [arXiv:astro-ph/0402453]}}.
\newblock {\url{https://doi.org/10.1086/422088}}.

\bibitem[{Mej{\'{\i}}a-Restrepo} \em{et~al.}(2016){Mej{\'{\i}}a-Restrepo},
  {Trakhtenbrot}, {Lira}, {Netzer}, and {Capellupo}]{mejia-restrepoetal16}
{Mej{\'{\i}}a-Restrepo}, J.E.; {Trakhtenbrot}, B.; {Lira}, P.; {Netzer}, H.;
  {Capellupo}, D.M.
\newblock {Active galactic nuclei at z\~{}1.5: II. Black Hole Mass estimation
  by means of broad emission lines}.
\newblock {\em \mnras} {\bf 2016}, {\em 460},
  \href{http://xxx.lanl.gov/abs/1603.03437}{{\normalfont [1603.03437]}}.

\bibitem[{Vietri} \em{et~al.}(2020){Vietri}, {Mainieri}, {Kakkad}, {Netzer},
  {Perna}, {Circosta}, {Harrison}, {Zappacosta}, {Husemann}, {Padovani},
  {Bischetti}, {Bongiorno}, {Brusa}, {Carniani}, {Cicone}, {Comastri},
  {Cresci}, {Feruglio}, {Fiore}, {Lanzuisi}, {Mannucci}, {Marconi},
  {Piconcelli}, {Puglisi}, {Salvato}, {Schramm}, {Schulze}, {Scholtz},
  {Vignali}, and {Zamorani}]{vietrietal20}
{Vietri}, G.; {Mainieri}, V.; {Kakkad}, D.; {Netzer}, H.; {Perna}, M.;
  {Circosta}, C.; {Harrison}, C.M.; {Zappacosta}, L.; {Husemann}, B.;
  {Padovani}, P.;  et~al.
\newblock {SUPER. III. Broad line region properties of AGNs at z
  {\ensuremath{\sim}} 2}.
\newblock {\em \aap} {\bf 2020}, {\em 644},~A175,
  \href{http://xxx.lanl.gov/abs/2010.07443}{{\normalfont
  [arXiv:astro-ph.GA/2010.07443]}}.
\newblock {\url{https://doi.org/10.1051/0004-6361/202039136}}.

\bibitem[{Wang} \em{et~al.}(2011){Wang}, {Wang}, {Zhou}, {Liu}, {Wang}, {Yuan},
  and {Dong}]{wangetal11}
{Wang}, H.; {Wang}, T.; {Zhou}, H.; {Liu}, B.; {Wang}, J.; {Yuan}, W.; {Dong},
  X.
\newblock {Coexistence of Gravitationally-bound and Radiation-driven C IV
  Emission Line Regions in Active Galactic Nuclei}.
\newblock {\em \apj} {\bf 2011}, {\em 738},~85,
  \href{http://xxx.lanl.gov/abs/1106.2584}{{\normalfont
  [arXiv:astro-ph.CO/1106.2584]}}.
\newblock {\url{https://doi.org/10.1088/0004-637X/738/1/85}}.

\bibitem[{Marconi} \em{et~al.}(2009){Marconi}, {Axon}, {Maiolino}, {Nagao},
  {Pietrini}, {Risaliti}, {Robinson}, and {Torricelli}]{marconietal09}
{Marconi}, A.; {Axon}, D.J.; {Maiolino}, R.; {Nagao}, T.; {Pietrini}, P.;
  {Risaliti}, G.; {Robinson}, A.; {Torricelli}, G.
\newblock {On the Observed Distributions of Black Hole Masses and Eddington
  Ratios from Radiation Pressure Corrected Virial Indicators}.
\newblock {\em \apjl} {\bf 2009}, {\em 698},~L103--L107,
  \href{http://xxx.lanl.gov/abs/0905.0539}{{\normalfont [0905.0539]}}.
\newblock {\url{https://doi.org/10.1088/0004-637X/698/2/L103}}.

\bibitem[{Wang} \em{et~al.}(2006){Wang}, {Wei}, and {He}]{wangetal06}
{Wang}, J.; {Wei}, J.Y.; {He}, X.T.
\newblock {A Sample of IRAS Infrared-selected Seyfert 1.5 Galaxies: Infrared
  Color {$\alpha$}(60, 25)-dominated Eigenvector 1}.
\newblock {\em \apj} {\bf 2006}, {\em 638},~106--119,
  \href{http://xxx.lanl.gov/abs/arXiv:astro-ph/0510564}{{\normalfont
  [arXiv:astro-ph/0510564]}}.
\newblock {\url{https://doi.org/10.1086/498667}}.

\bibitem[{Wolf} \em{et~al.}(2020){Wolf}, {Salvato}, {Coffey}, {Merloni},
  {Buchner}, {Arcodia}, {Baron}, {Carrera}, {Comparat}, {Schneider}, and
  {Nandra}]{wolfetal20}
{Wolf}, J.; {Salvato}, M.; {Coffey}, D.; {Merloni}, A.; {Buchner}, J.;
  {Arcodia}, R.; {Baron}, D.; {Carrera}, F.J.; {Comparat}, J.; {Schneider},
  D.P.;  et~al.
\newblock {Exploring the diversity of Type 1 active galactic nuclei identified
  in SDSS-IV/SPIDERS}.
\newblock {\em \mnras} {\bf 2020}, {\em 492},~3580--3601,
  \href{http://xxx.lanl.gov/abs/1911.01947}{{\normalfont
  [arXiv:astro-ph.HE/1911.01947]}}.
\newblock {\url{https://doi.org/10.1093/mnras/staa018}}.

\bibitem[{Peterson} and {Ferland}(1986)]{petersonferland86}
{Peterson}, B.M.; {Ferland}, G.J.
\newblock {An accretion event in the Seyfert galaxy NGC 5548}.
\newblock {\em \nat} {\bf 1986}, {\em 324},~345--347.
\newblock {\url{https://doi.org/10.1038/324345a0}}.

\bibitem[{Snedden} and {Gaskell}(2004)]{sneddengaskell04}
{Snedden}, S.; {Gaskell}, C.
\newblock {Different Velocity Dependences of Physical Conditions of High- and
  Low-Ionization Lines in Broad-Line Regions}.
\newblock In Proceedings of the AGN Physics with the Sloan Digital Sky Survey;
  {Richards}, G.T.; {Hall}, P.B., Eds.,  2004, Vol. 311, {\em Astronomical
  Society of the Pacific Conference Series}, p. 197,
  \href{http://xxx.lanl.gov/abs/astro-ph/0403174}{{\normalfont
  [astro-ph/0403174]}}.

\bibitem[{Morris} and {Ward}(1989)]{morrisward89}
{Morris}, S.L.; {Ward}, M.J.
\newblock {Optically thin gas in the broad-line region of Seyfert galaxies}.
\newblock {\em ApJ} {\bf 1989}, {\em 340},~713--728.
\newblock {\url{https://doi.org/10.1086/167432}}.

\bibitem[{Netzer}(1977)]{netzer77}
{Netzer}, H.
\newblock {On the profiles of the broad lines in the spectra of QSOs and
  Seyfert galaxies}.
\newblock {\em \mnras} {\bf 1977}, {\em 181},~89P--92P.

\bibitem[{Corbin}(1995)]{corbin95}
{Corbin}, M.R.
\newblock {QSO Broad Emission Line Asymmetries: Evidence of Gravitational
  Redshift?}
\newblock {\em \apj} {\bf 1995}, {\em 447},~496.
\newblock {\url{https://doi.org/10.1086/175894}}.

\bibitem[{Popovic} \em{et~al.}(1995){Popovic}, {Vince},
  {Atanackovic-Vukmanovic}, and {Kubicela}]{popovicetal95}
{Popovic}, L.C.; {Vince}, I.; {Atanackovic-Vukmanovic}, O.; {Kubicela}, A.
\newblock {Contribution of gravitational redshift to spectral line profiles of
  Seyfert galaxies and quasars.}
\newblock {\em \aap} {\bf 1995}, {\em 293},~309--314.

\bibitem[{Mu{\~n}oz} \em{et~al.}(2003){Mu{\~n}oz}, {Falco}, {Kochanek},
  {Leh{\'a}r}, and {Mediavilla}]{munozetal03}
{Mu{\~n}oz}, J.A.; {Falco}, E.E.; {Kochanek}, C.S.; {Leh{\'a}r}, J.;
  {Mediavilla}, E.
\newblock {The Redshift Distribution of Flat-Spectrum Radio Sources}.
\newblock {\em \apj} {\bf 2003}, {\em 594},~684--694,
  \href{http://xxx.lanl.gov/abs/astro-ph/0305382}{{\normalfont
  [arXiv:astro-ph/astro-ph/0305382]}}.
\newblock {\url{https://doi.org/10.1086/377077}}.

\bibitem[{Mediavilla} \em{et~al.}(2018){Mediavilla}, {Jim{\'e}nez-Vicente},
  {Fian}, {Mu{\~n}oz}, {Falco}, {Motta}, and {Guerras}]{mediavillaetal18}
{Mediavilla}, E.; {Jim{\'e}nez-Vicente}, J.; {Fian}, C.; {Mu{\~n}oz}, J.A.;
  {Falco}, E.; {Motta}, V.; {Guerras}, E.
\newblock {Systematic Redshift of the Fe III UV Lines in Quasars: Measuring
  Supermassive Black Hole Masses under the Gravitational Redshift Hypothesis}.
\newblock {\em \apj} {\bf 2018}, {\em 862},~104,
  \href{http://xxx.lanl.gov/abs/1807.04048}{{\normalfont
  [arXiv:astro-ph.GA/1807.04048]}}.
\newblock {\url{https://doi.org/10.3847/1538-4357/aacbd3}}.

\bibitem[{Fian} \em{et~al.}(2022){Fian}, {Mediavilla}, {Jim{\'e}nez-Vicente},
  {Motta}, {Mu{\~n}oz}, {Chelouche}, and {Hanslmeier}]{fianetal22}
{Fian}, C.; {Mediavilla}, E.; {Jim{\'e}nez-Vicente}, J.; {Motta}, V.;
  {Mu{\~n}oz}, J.A.; {Chelouche}, D.; {Hanslmeier}, A.
\newblock {Revealing the structure of the lensed quasar Q 0957+561. II.
  Supermassive black hole mass via gravitational redshift}.
\newblock {\em \aap} {\bf 2022}, {\em 667},~A67,
  \href{http://xxx.lanl.gov/abs/2107.11791}{{\normalfont
  [arXiv:astro-ph.GA/2107.11791]}}.
\newblock {\url{https://doi.org/10.1051/0004-6361/202140977}}.

\bibitem[{Wang} \em{et~al.}(2014){Wang}, {Du}, {Hu}, {Netzer}, {Bai}, {Lu},
  {Kaspi}, {Qiu}, {Li}, {Wang}, and {SEAMBH Collaboration}]{wangetal14b}
{Wang}, J.M.; {Du}, P.; {Hu}, C.; {Netzer}, H.; {Bai}, J.M.; {Lu}, K.X.;
  {Kaspi}, S.; {Qiu}, J.; {Li}, Y.R.; {Wang}, F.;  et~al.
\newblock {Supermassive Black Holes with High Accretion Rates in Active
  Galactic Nuclei. II. The Most Luminous Standard Candles in the Universe}.
\newblock {\em \apj} {\bf 2014}, {\em 793},~108,
  \href{http://xxx.lanl.gov/abs/1408.2337}{{\normalfont
  [arXiv:astro-ph.HE/1408.2337]}}.
\newblock {\url{https://doi.org/10.1088/0004-637X/793/2/108}}.

\bibitem[Marziani(2023)]{marziani23a}
Marziani, P.
\newblock Accretion/Ejection Phenomena and Emission-Line Profile (A)symmetries
  in Type-1 Active Galactic Nuclei.
\newblock {\em Symmetry} {\bf 2023}, {\em 15}.
\newblock {\url{https://doi.org/10.3390/sym15101859}}.

\bibitem[{Pagel} \em{et~al.}(1979){Pagel}, {Edmunds}, {Blackwell}, {Chun}, and
  {Smith}]{pageletal79}
{Pagel}, B.E.J.; {Edmunds}, M.G.; {Blackwell}, D.E.; {Chun}, M.S.; {Smith}, G.
\newblock {On the composition of H II regions in southern galaxies - I. NGC 300
  and 1365.}
\newblock {\em \mnras} {\bf 1979}, {\em 189},~95--113.
\newblock {\url{https://doi.org/10.1093/mnras/189.1.95}}.

\bibitem[{Huang} \em{et~al.}(2023){Huang}, {Lin}, and {Shields}]{huangetal23}
{Huang}, J.; {Lin}, D.N.C.; {Shields}, G.
\newblock {Metal enrichment due to embedded stars in AGN discs}.
\newblock {\em \mnras} {\bf 2023},
  \href{http://xxx.lanl.gov/abs/2308.15761}{{\normalfont
  [arXiv:astro-ph.GA/2308.15761]}}.
\newblock {\url{https://doi.org/10.1093/mnras/stad2642}}.

\bibitem[{Ferland} \em{et~al.}(1996){Ferland}, {Baldwin}, {Korista}, {Hamann},
  {Carswell}, {Phillips}, {Wilkes}, and {Williams}]{ferlandetal96}
{Ferland}, G.J.; {Baldwin}, J.A.; {Korista}, K.T.; {Hamann}, F.; {Carswell},
  R.F.; {Phillips}, M.; {Wilkes}, B.; {Williams}, R.E.
\newblock {High Metal Enrichments in Luminous Quasars}.
\newblock {\em \apj} {\bf 1996}, {\em 461},~683--+.
\newblock {\url{https://doi.org/10.1086/177094}}.

\bibitem[{Hamann} \em{et~al.}(2002){Hamann}, {Korista}, {Ferland}, {Warner},
  and {Baldwin}]{hamannetal02}
{Hamann}, F.; {Korista}, K.T.; {Ferland}, G.J.; {Warner}, C.; {Baldwin}, J.
\newblock {Metallicities and Abundance Ratios from Quasar Broad Emission
  Lines}.
\newblock {\em \apj} {\bf 2002}, {\em 564},~592--603,
  \href{http://xxx.lanl.gov/abs/astro-ph/0109006}{{\normalfont
  [arXiv:astro-ph/astro-ph/0109006]}}.
\newblock {\url{https://doi.org/10.1086/324289}}.

\bibitem[{Osmer} and {Smith}(1976)]{osmersmith76}
{Osmer}, P.S.; {Smith}, M.G.
\newblock {The emission-line spectra of nine newly discovered, optically
  selected quasars with redshift 2.5 to 3.1}.
\newblock {\em \apj} {\bf 1976}, {\em 210},~267--276.
\newblock {\url{https://doi.org/10.1086/154827}}.

\bibitem[{Shields}(1976)]{shields76}
{Shields}, G.A.
\newblock {The abundance of nitrogen in QSOs.}
\newblock {\em \apj} {\bf 1976}, {\em 204},~330--336.
\newblock {\url{https://doi.org/10.1086/154176}}.

\bibitem[{Vila-Costas} and {Edmunds}(1993)]{vila-costasedmunds93}
{Vila-Costas}, M.B.; {Edmunds}, M.G.
\newblock {The nitrogen-to-oxygen ratio in galaxies and its implications for
  the origin of nitrogen.}
\newblock {\em \mnras} {\bf 1993}, {\em 265},~199--212.
\newblock {\url{https://doi.org/10.1093/mnras/265.1.199}}.

\bibitem[{Izotov} and {Thuan}(1999)]{izotovthuan99}
{Izotov}, Y.I.; {Thuan}, T.X.
\newblock {Heavy-Element Abundances in Blue Compact Galaxies}.
\newblock {\em \apj} {\bf 1999}, {\em 511},~639--659,
  \href{http://xxx.lanl.gov/abs/astro-ph/9811387}{{\normalfont
  [arXiv:astro-ph/astro-ph/9811387]}}.
\newblock {\url{https://doi.org/10.1086/306708}}.

\bibitem[{Collin} and {Joly}(2000)]{collinjoly00}
{Collin}, S.; {Joly}, M.
\newblock {The Fe II problem in NLS1s}.
\newblock {\em NAR} {\bf 2000}, {\em 44},~531--537,
  \href{http://xxx.lanl.gov/abs/arXiv:astro-ph/0005153}{{\normalfont
  [arXiv:astro-ph/0005153]}}.
\newblock {\url{https://doi.org/10.1016/S1387-6473(00)00093-2}}.

\bibitem[{Matsuoka} \em{et~al.}(2008){Matsuoka}, {Kawara}, and
  {Oyabu}]{matsuokaetal08}
{Matsuoka}, Y.; {Kawara}, K.; {Oyabu}, S.
\newblock {Low-Ionization Emission Regions in Quasars: Gas Properties Probed
  with Broad O I and Ca II Lines}.
\newblock {\em ApJ} {\bf 2008}, {\em 673},~62--68,
  \href{http://xxx.lanl.gov/abs/0710.2954}{{\normalfont [0710.2954]}}.
\newblock {\url{https://doi.org/10.1086/524193}}.

\bibitem[{Ferland} \em{et~al.}(2017){Ferland}, {Chatzikos}, {Guzm{\'a}n},
  {Lykins}, {van Hoof}, {Williams}, {Abel}, {Badnell}, {Keenan}, {Porter}, and
  {Stancil}]{ferlandetal17}
{Ferland}, G.J.; {Chatzikos}, M.; {Guzm{\'a}n}, F.; {Lykins}, M.L.; {van Hoof},
  P.A.M.; {Williams}, R.J.R.; {Abel}, N.P.; {Badnell}, N.R.; {Keenan}, F.P.;
  {Porter}, R.L.;  et~al.
\newblock {The 2017 Release Cloudy}.
\newblock {\em \rmxaa} {\bf 2017}, {\em 53},~385--438,
  \href{http://xxx.lanl.gov/abs/1705.10877}{{\normalfont
  [arXiv:astro-ph.GA/1705.10877]}}.

\bibitem[{Baldwin} \em{et~al.}(1995){Baldwin}, {Ferland}, {Korista}, and
  {Verner}]{baldwinetal95}
{Baldwin}, J.; {Ferland}, G.; {Korista}, K.; {Verner}, D.
\newblock {Locally Optimally Emitting Clouds and the Origin of Quasar Emission
  Lines}.
\newblock {\em ApJL} {\bf 1995}, {\em 455},~L119+,
  \href{http://xxx.lanl.gov/abs/arXiv:astro-ph/9510080}{{\normalfont
  [arXiv:astro-ph/9510080]}}.
\newblock {\url{https://doi.org/10.1086/309827}}.

\bibitem[{Korista} \em{et~al.}(1997){Korista}, {Baldwin}, {Ferland}, and
  {Verner}]{koristaetal97}
{Korista}, K.; {Baldwin}, J.; {Ferland}, G.; {Verner}, D.
\newblock {An Atlas of Computed Equivalent Widths of Quasar Broad Emission
  Lines}.
\newblock {\em ApJS} {\bf 1997}, {\em 108},~401--+,
  \href{http://xxx.lanl.gov/abs/arXiv:astro-ph/9611220}{{\normalfont
  [arXiv:astro-ph/9611220]}}.
\newblock {\url{https://doi.org/10.1086/312966}}.

\bibitem[{Mathews} and {Ferland}(1987)]{mathewsferland87}
{Mathews}, W.G.; {Ferland}, G.J.
\newblock {What heats the hot phase in active nuclei?}
\newblock {\em ApJ} {\bf 1987}, {\em 323},~456--467.
\newblock {\url{https://doi.org/10.1086/165843}}.

\bibitem[{Ferland} \em{et~al.}(2020){Ferland}, {Done}, {Jin}, {Landt}, and
  {Ward}]{ferlandetal20}
{Ferland}, G.J.; {Done}, C.; {Jin}, C.; {Landt}, H.; {Ward}, M.J.
\newblock {State-of-the-art AGN SEDs for photoionization models: BLR
  predictions confront the observations}.
\newblock {\em \mnras} {\bf 2020}, {\em 494},~5917--5922,
  \href{http://xxx.lanl.gov/abs/2004.11873}{{\normalfont
  [arXiv:astro-ph.HE/2004.11873]}}.
\newblock {\url{https://doi.org/10.1093/mnras/staa1207}}.

\bibitem[{Sigut} and {Pradhan}(1998)]{sigutpradhan98}
{Sigut}, T.A.A.; {Pradhan}, A.K.
\newblock {Ly alpha Flourescent Excitation of Fe II in Active Galactic Nuclei}.
\newblock {\em \apjl} {\bf 1998}, {\em 499},~L139,
  \href{http://xxx.lanl.gov/abs/arXiv:astro-ph/9804183}{{\normalfont
  [arXiv:astro-ph/9804183]}}.
\newblock {\url{https://doi.org/10.1086/311369}}.

\bibitem[{Sigut} and {Pradhan}(2003)]{sigutpradhan03}
{Sigut}, T.A.A.; {Pradhan}, A.K.
\newblock {Predicted Fe II Emission-Line Strengths from Active Galactic
  Nuclei}.
\newblock {\em ApJS} {\bf 2003}, {\em 145},~15--37,
  \href{http://xxx.lanl.gov/abs/arXiv:astro-ph/0206096}{{\normalfont
  [arXiv:astro-ph/0206096]}}.
\newblock {\url{https://doi.org/10.1086/345498}}.

\bibitem[D'Agostini(2003)]{dagostini03}
D'Agostini, G.
\newblock {\em {Bayesian Reasoning in Data Analysis: A Critical Introduction}};
  World Scientific: Singapore,  2003.
\newblock {\url{https://doi.org/10.1142/5262}}.

\bibitem[{Nagao} \em{et~al.}(2006){Nagao}, {Maiolino}, and
  {Marconi}]{nagaoetal06b}
{Nagao}, T.; {Maiolino}, R.; {Marconi}, A.
\newblock {Gas metallicity in the narrow-line regions of high-redshift active
  galactic nuclei}.
\newblock {\em \aap} {\bf 2006}, {\em 447},~863--876,
  \href{http://xxx.lanl.gov/abs/arXiv:astro-ph/0508652}{{\normalfont
  [arXiv:astro-ph/0508652]}}.
\newblock {\url{https://doi.org/10.1051/00\-04-63\-61\-:20\-054127}}.

\bibitem[{Maiolino} and {Mannucci}(2019)]{maiolinomannucci19}
{Maiolino}, R.; {Mannucci}, F.
\newblock {De re metallica: the cosmic chemical evolution of galaxies}.
\newblock {\em The Astronomy and Astrophysics Review} {\bf 2019}, {\em 27},~3,
  \href{http://xxx.lanl.gov/abs/1811.09642}{{\normalfont
  [arXiv:astro-ph.GA/1811.09642]}}.
\newblock {\url{https://doi.org/10.1007/s00159-018-0112-2}}.

\bibitem[{Wang} \em{et~al.}(2023){Wang}, {Songsheng}, {Li}, and
  {Du}]{wangetal23}
{Wang}, J.M.; {Songsheng}, Y.Y.; {Li}, Y.R.; {Du}, P.
\newblock {Final stage of merging binaries of supermassive black holes:
  observational signatures}.
\newblock {\em \mnras} {\bf 2023}, {\em 518},~3397--3406,
  \href{http://xxx.lanl.gov/abs/2211.03947}{{\normalfont
  [arXiv:astro-ph.GA/2211.03947]}}.
\newblock {\url{https://doi.org/10.1093/mnras/stac3266}}.

\bibitem[{Cantiello} \em{et~al.}(2021){Cantiello}, {Jermyn}, and
  {Lin}]{cantielloetal21}
{Cantiello}, M.; {Jermyn}, A.S.; {Lin}, D.N.C.
\newblock {Stellar Evolution in AGN Disks}.
\newblock {\em \apj} {\bf 2021}, {\em 910},~94,
  \href{http://xxx.lanl.gov/abs/2009.03936}{{\normalfont
  [arXiv:astro-ph.SR/2009.03936]}}.
\newblock {\url{https://doi.org/10.3847/1538-4357/abdf4f}}.

\bibitem[{Heckman} \em{et~al.}(1989){Heckman}, {Baum}, {van Breugel}, and
  {McCarthy}]{heckmanetal89}
{Heckman}, T.M.; {Baum}, S.A.; {van Breugel}, W.J.M.; {McCarthy}, P.
\newblock {Dynamical, Physical, and Chemical Properties of Emission-Line
  Nebulae in Cooling Flows}.
\newblock {\em \apj} {\bf 1989}, {\em 338},~48.
\newblock {\url{https://doi.org/10.1086/167181}}.

\bibitem[{Lim} \em{et~al.}(2008){Lim}, {Ao}, and {Dinh-V-Trung}]{limetal08}
{Lim}, J.; {Ao}, Y.; {Dinh-V-Trung}.
\newblock {Radially Inflowing Molecular Gas in NGC 1275 Deposited by an X-Ray
  Cooling Flow in the Perseus Cluster}.
\newblock {\em \apj} {\bf 2008}, {\em 672},~252--265,
  \href{http://xxx.lanl.gov/abs/0712.2979}{{\normalfont
  [arXiv:astro-ph/0712.2979]}}.
\newblock {\url{https://doi.org/10.1086/523664}}.

\bibitem[{Sun} and {Malkan}(1989)]{sunmalkan89}
{Sun}, W.H.; {Malkan}, M.A.
\newblock {Fitting Improved Accretion Disk Models to the Multiwavelength
  Continua of Quasars and Active Galactic Nuclei}.
\newblock {\em \apj} {\bf 1989}, {\em 346},~68.
\newblock {\url{https://doi.org/10.1086/167986}}.

\bibitem[{Sulentic} \em{et~al.}(2007){Sulentic}, {Bachev}, {Marziani},
  {Negrete}, and {Dultzin}]{sulenticetal07}
{Sulentic}, J.W.; {Bachev}, R.; {Marziani}, P.; {Negrete}, C.A.; {Dultzin}, D.
\newblock {C IV {\ensuremath{\lambda}}1549 as an Eigenvector 1 Parameter for
  Active Galactic Nuclei}.
\newblock {\em \apj} {\bf 2007}, {\em 666},~757--777,
  \href{http://xxx.lanl.gov/abs/0705.1895}{{\normalfont
  [arXiv:astro-ph/0705.1895]}}.
\newblock {\url{https://doi.org/10.1086/519916}}.

\bibitem[{Mauch} \em{et~al.}(2003){Mauch}, {Murphy}, {Buttery}, {Curran},
  {Hunstead}, {Piestrzynski}, {Robertson}, and {Sadler}]{mauchetal03}
{Mauch}, T.; {Murphy}, T.; {Buttery}, H.J.; {Curran}, J.; {Hunstead}, R.W.;
  {Piestrzynski}, B.; {Robertson}, J.G.; {Sadler}, E.M.
\newblock {SUMSS: a wide-field radio imaging survey of the southern sky - II.
  The source catalogue}.
\newblock {\em \mnras} {\bf 2003}, {\em 342},~1117--1130,
  \href{http://xxx.lanl.gov/abs/astro-ph/0303188}{{\normalfont
  [arXiv:astro-ph/astro-ph/0303188]}}.
\newblock {\url{https://doi.org/10.1046/j.1365-8711.2003.06605.x}}.

\bibitem[{Sikora} \em{et~al.}(2007){Sikora}, {Stawarz}, and
  {Lasota}]{sikoraetal07}
{Sikora}, M.; {Stawarz}, {\L}.; {Lasota}, J.P.
\newblock {Radio Loudness of Active Galactic Nuclei: Observational Facts and
  Theoretical Implications}.
\newblock {\em \apj} {\bf 2007}, {\em 658},~815--828,
  \href{http://xxx.lanl.gov/abs/astro-ph/0604095}{{\normalfont
  [astro-ph/0604095]}}.
\newblock {\url{https://doi.org/10.1086/511972}}.

\bibitem[{Peterson} \em{et~al.}(2004){Peterson}, {Ferrarese}, {Gilbert},
  {Kaspi}, {Malkan}, {Maoz}, {Merritt}, {Netzer}, {Onken}, {Pogge},
  {Vestergaard}, and {Wandel}]{petersonetal04}
{Peterson}, B.M.; {Ferrarese}, L.; {Gilbert}, K.M.; {Kaspi}, S.; {Malkan},
  M.A.; {Maoz}, D.; {Merritt}, D.; {Netzer}, H.; {Onken}, C.A.; {Pogge}, R.W.;
  et~al.
\newblock {Central Masses and Broad-Line Region Sizes of Active Galactic
  Nuclei. II. A Homogeneous Analysis of a Large Reverberation-Mapping
  Database}.
\newblock {\em ApJ} {\bf 2004}, {\em 613},~682--699,
  \href{http://xxx.lanl.gov/abs/arXiv:astro-ph/0407299}{{\normalfont
  [arXiv:astro-ph/0407299]}}.
\newblock {\url{https://doi.org/10.1086/423269}}.

\bibitem[{Bentz} and {Katz}(2015)]{bentzkatz15}
{Bentz}, M.C.; {Katz}, S.
\newblock {The AGN Black Hole Mass Database}.
\newblock {\em Publications of the Astronomical Society of the Pacific} {\bf
  2015}, {\em 127},~67,  \href{http://xxx.lanl.gov/abs/1411.2596}{{\normalfont
  [arXiv:astro-ph.GA/1411.2596]}}.
\newblock {\url{https://doi.org/10.1086/679601}}.

\bibitem[{Jiang} \em{et~al.}(2021){Jiang}, {Marziani}, {Savi{\'c}},
  {Shablovinskaya}, {Popovi{\'c}}, {Afanasiev}, {Czerny}, {Wang}, {del Olmo},
  {D'Onofrio}, {{\'S}niegowska}, {Mazzei}, and {Panda}]{jiangetal21}
{Jiang}, B.W.; {Marziani}, P.; {Savi{\'c}}, {\DJ}.; {Shablovinskaya}, E.;
  {Popovi{\'c}}, L.{\v{C}}.; {Afanasiev}, V.L.; {Czerny}, B.; {Wang}, J.M.;
  {del Olmo}, A.; {D'Onofrio}, M.;  et~al.
\newblock {Linear Spectropolarimetric Analysis of Fairall 9 with VLT/FORS2}.
\newblock {\em \mnras} {\bf 2021},
  \href{http://xxx.lanl.gov/abs/2108.00983}{{\normalfont
  [arXiv:astro-ph.GA/2108.00983]}}.
\newblock {\url{https://doi.org/10.1093/mnras/stab2273}}.

\bibitem[{Buendia-Rios} \em{et~al.}(2023){Buendia-Rios}, {Negrete}, {Marziani},
  and {Dultzin}]{buendia-riosetal23}
{Buendia-Rios}, T.M.; {Negrete}, C.A.; {Marziani}, P.; {Dultzin}, D.
\newblock {Statistical analysis of Al III and C III] emission lines as virial
  black hole mass estimators in quasars}.
\newblock {\em \aap} {\bf 2023}, {\em 669},~A135,
  \href{http://xxx.lanl.gov/abs/2209.05526}{{\normalfont
  [arXiv:astro-ph.GA/2209.05526]}}.
\newblock {\url{https://doi.org/10.1051/0004-6361/202244177}}.

\bibitem[Azzalini and Regoli(2012)]{azzaliniregoli12}
Azzalini, A.; Regoli, G.
\newblock Some properties of skew-symmetric distributions.
\newblock {\em Ann. Inst. Statist. Math.} {\bf 2012}, {\em 64},~857--879.
\newblock {\url{https://doi.org/10.1007/s10463-011-0338-5}}.

\bibitem[{Marshall} \em{et~al.}(1996){Marshall}, {Carone}, {Shull}, {Malkan},
  and {Elvis}]{marshalletal96}
{Marshall}, H.L.; {Carone}, T.E.; {Shull}, J.M.; {Malkan}, M.A.; {Elvis}, M.
\newblock {The Steep Soft X-Ray Spectrum of the Highly Variable Active Nucleus
  in Markarian 478}.
\newblock {\em \apj} {\bf 1996}, {\em 457},~169.
\newblock {\url{https://doi.org/10.1086/176720}}.

\bibitem[{Hwang} and {Bowyer}(1997)]{hwangetal97}
{Hwang}, C.Y.; {Bowyer}, S.
\newblock {The Extreme-Ultraviolet Emission of the Seyfert Galaxies Markarian
  279, Markarian 478, and Ton S180}.
\newblock {\em \apj} {\bf 1997}, {\em 475},~552--556.
\newblock {\url{https://doi.org/10.1086/303562}}.

\bibitem[{Yuan} \em{et~al.}(2004){Yuan}, {Brotherton}, {Green}, and
  {Kriss}]{yuanetal04}
{Yuan}, Q.; {Brotherton}, M.; {Green}, R.F.; {Kriss}, G.A.
\newblock {Outflowing Components in the Prototype Narrow-Line Seyfert 1 Galaxy
  Markarian 478}.
\newblock In Proceedings of the Recycling Intergalactic and Interstellar
  Matter; {Duc}, P.A.; {Braine}, J.; {Brinks}, E., Eds.,  2004, Vol. 217, p.
  364.

\bibitem[{Zacharias} \em{et~al.}(2005){Zacharias}, {Monet}, {Levine}, {Urban},
  {Gaume}, and {Wycoff}]{zachariasetal05}
{Zacharias}, N.; {Monet}, D.G.; {Levine}, S.E.; {Urban}, S.E.; {Gaume}, R.;
  {Wycoff}, G.L.
\newblock {VizieR Online Data Catalog: NOMAD Catalog (Zacharias+ 2005)}.
\newblock {\em VizieR Online Data Catalog} {\bf 2005}, {\em 1297}.

\bibitem[{Vestergaard}(2006)]{vestergaard06}
{Vestergaard}, M.
\newblock {A first step toward constraining supermassive black-hole growth}.
\newblock {\em \nar} {\bf 2006}, {\em 50},~817--820.
\newblock {\url{https://doi.org/10.1016/j.newar.2006.06.039}}.

\bibitem[{Kellermann} \em{et~al.}(1989){Kellermann}, {Sramek}, {Schmidt},
  {Shaffer}, and {Green}]{kellermannetal89}
{Kellermann}, K.I.; {Sramek}, R.; {Schmidt}, M.; {Shaffer}, D.B.; {Green}, R.
\newblock {VLA observations of objects in the Palomar Bright Quasar Survey}.
\newblock {\em \aj} {\bf 1989}, {\em 98},~1195--1207.
\newblock {\url{https://doi.org/10.1086/115207}}.

\bibitem[{Popovi{\'c}} \em{et~al.}(2004){Popovi{\'c}}, {Mediavilla}, {Bon}, and
  {Ili{\'c}}]{popovicetal04}
{Popovi{\'c}}, L.{\v{C}}.; {Mediavilla}, E.; {Bon}, E.; {Ili{\'c}}, D.
\newblock {Contribution of the disk emission to the broad emission lines in
  AGNs: Two-component model}.
\newblock {\em \aap} {\bf 2004}, {\em 423},~909--918,
  \href{http://xxx.lanl.gov/abs/astro-ph/0405447}{{\normalfont
  [arXiv:astro-ph/astro-ph/0405447]}}.
\newblock {\url{https://doi.org/10.1051/0004-6361:20034431}}.

\bibitem[{Snedden} and {Gaskell}(2007)]{sneddengaskell07}
{Snedden}, S.A.; {Gaskell}, C.M.
\newblock {The Case for Optically Thick High-Velocity Broad-Line Region Gas in
  Active Galactic Nuclei}.
\newblock {\em ApJ} {\bf 2007}, {\em 669},~126--134.
\newblock {\url{https://doi.org/10.1086/521290}}.

\bibitem[{Bon} \em{et~al.}(2009){Bon}, {Gavrilovic}, {La Mura}, and
  {Popovic}]{bonetal09}
{Bon}, E.; {Gavrilovic}, N.; {La Mura}, G.; {Popovic}, L.C.
\newblock {Complex Broad Emission Line Profiles of AGN - Geometry of the Broad
  Line Region}.
\newblock {\em ArXiv e-prints} {\bf 2009},
  \href{http://xxx.lanl.gov/abs/0910.0991}{{\normalfont [0910.0991]}}.

\bibitem[{Vietri}(2017)]{vietrietal17}
{Vietri}, G.
\newblock {The LBT/WISSH quasar survey: revealing powerful winds in the most
  luminous AGN}.
\newblock In Proceedings of the American Astronomical Society Meeting
  Abstracts,  2017, Vol. 229, {\em American Astronomical Society Meeting
  Abstracts}, p. 302.06.

\bibitem[{Yang} \em{et~al.}(2023){Yang}, {Wang}, {Fan}, {Hennawi}, {Barth},
  {Ba{\~n}ados}, {Sun}, {Liu}, {Cai}, {Jiang}, {Li}, {Onoue}, {Schindler},
  {Shen}, {Wu}, {Bhowmick}, {Bieri}, {Blecha}, {Bosman}, {Champagne}, {Colina},
  {Connor}, {Costa}, {Davies}, {Decarli}, {De Rosa}, {Drake}, {Egami},
  {Eilers}, {Evans}, {Farina}, {Habouzit}, {Haiman}, {Jin}, {Jun}, {Kakiichi},
  {Khusanova}, {Kulkarni}, {Loiacono}, {Lupi}, {Mazzucchelli}, {Pan},
  {Rojas-Ruiz}, {Strauss}, {Tee}, {Trakhtenbrot}, {Trebitsch}, {Venemans},
  {Vestergaard}, {Volonteri}, {Walter}, {Xie}, {Yue}, {Zhang}, {Zhang}, and
  {Zou}]{yangetal23}
{Yang}, J.; {Wang}, F.; {Fan}, X.; {Hennawi}, J.F.; {Barth}, A.J.;
  {Ba{\~n}ados}, E.; {Sun}, F.; {Liu}, W.; {Cai}, Z.; {Jiang}, L.;  et~al.
\newblock {A SPectroscopic Survey of Biased Halos in the Reionization Era
  (ASPIRE): A First Look at the Rest-frame Optical Spectra of z > 6.5 Quasars
  Using JWST}.
\newblock {\em \apjl} {\bf 2023}, {\em 951},~L5,
  \href{http://xxx.lanl.gov/abs/2304.09888}{{\normalfont
  [arXiv:astro-ph.GA/2304.09888]}}.
\newblock {\url{https://doi.org/10.3847/2041-8213/acc9c8}}.

\bibitem[{Shang} \em{et~al.}(2007){Shang}, {Wills}, {Wills}, and
  {Brotherton}]{shangetal07}
{Shang}, Z.; {Wills}, B.J.; {Wills}, D.; {Brotherton}, M.S.
\newblock {Spectral Properties from Ly{$\alpha$} to H{$\alpha$} for an
  Essentially Complete Sample of Quasars. I. Data}.
\newblock {\em AJ} {\bf 2007}, {\em 134},~294--393,
  \href{http://xxx.lanl.gov/abs/arXiv:astro-ph/0703690}{{\normalfont
  [arXiv:astro-ph/0703690]}}.
\newblock {\url{https://doi.org/10.1086/518505}}.

\bibitem[{Marziani} \em{et~al.}(2013){Marziani}, {Sulentic}, {Plauchu-Frayn},
  and {del Olmo}]{marzianietal13a}
{Marziani}, P.; {Sulentic}, J.W.; {Plauchu-Frayn}, I.; {del Olmo}, A.
\newblock {Is Mg II 2800 a Reliable Virial Broadening Estimator for Quasars?}
\newblock {\em AAp} {\bf 2013}, {\em 555},~89, 16pp,
  \href{http://xxx.lanl.gov/abs/1305.1096}{{\normalfont
  [arXiv:astro-ph.CO/1305.1096]}}.

\bibitem[{Feruglio} \em{et~al.}(2010){Feruglio}, {Maiolino}, {Piconcelli},
  {Menci}, {Aussel}, {Lamastra}, and {Fiore}]{feruglioetal10}
{Feruglio}, C.; {Maiolino}, R.; {Piconcelli}, E.; {Menci}, N.; {Aussel}, H.;
  {Lamastra}, A.; {Fiore}, F.
\newblock {Quasar feedback revealed by giant molecular outflows}.
\newblock {\em \aap} {\bf 2010}, {\em 518},~L155,
  \href{http://xxx.lanl.gov/abs/1006.1655}{{\normalfont
  [arXiv:astro-ph.CO/1006.1655]}}.
\newblock {\url{https://doi.org/10.1051/0004-6361/201015164}}.

\bibitem[{Harrison} \em{et~al.}(2014){Harrison}, {Alexander}, {Mullaney}, and
  {Swinbank}]{harrisonetal14}
{Harrison}, C.M.; {Alexander}, D.M.; {Mullaney}, J.R.; {Swinbank}, A.M.
\newblock {Kiloparsec-scale outflows are prevalent among luminous AGN: outflows
  and feedback in the context of the overall AGN population}.
\newblock {\em \mnras} {\bf 2014}, {\em 441},~3306--3347,
  \href{http://xxx.lanl.gov/abs/1403.3086}{{\normalfont
  [arXiv:astro-ph.GA/1403.3086]}}.
\newblock {\url{https://doi.org/10.1093/mnras/stu515}}.

\bibitem[{Feruglio} \em{et~al.}(2015){Feruglio}, {Fiore}, {Carniani},
  {Piconcelli}, {Zappacosta}, {Bongiorno}, {Cicone}, {Maiolino}, {Marconi},
  {Menci}, {Puccetti}, and {Veilleux}]{feruglioetal15}
{Feruglio}, C.; {Fiore}, F.; {Carniani}, S.; {Piconcelli}, E.; {Zappacosta},
  L.; {Bongiorno}, A.; {Cicone}, C.; {Maiolino}, R.; {Marconi}, A.; {Menci},
  N.;  et~al.
\newblock {The multi-phase winds of Markarian 231: from the hot, nuclear,
  ultra-fast wind to the galaxy-scale, molecular outflow}.
\newblock {\em \aap} {\bf 2015}, {\em 583},~A99,
  \href{http://xxx.lanl.gov/abs/1503.01481}{{\normalfont
  [arXiv:astro-ph.GA/1503.01481]}}.
\newblock {\url{https://doi.org/10.1051/0004-6361/201526020}}.

\bibitem[{Woo} \em{et~al.}(2016){Woo}, {Bae}, {Son}, and {Karouzos}]{wooetal16}
{Woo}, J.H.; {Bae}, H.J.; {Son}, D.; {Karouzos}, M.
\newblock {The Prevalence of Gas Outflows in Type 2 AGNs}.
\newblock {\em \apj} {\bf 2016}, {\em 817},~108,
  \href{http://xxx.lanl.gov/abs/1511.05142}{{\normalfont
  [arXiv:astro-ph.GA/1511.05142]}}.
\newblock {\url{https://doi.org/10.3847/0004-637X/817/2/108}}.

\bibitem[{Kova{\v{c}}evi{\'c}-Doj{\v{c}}inovi{\'c}}
  \em{et~al.}(2022){Kova{\v{c}}evi{\'c}-Doj{\v{c}}inovi{\'c}},
  {Doj{\v{c}}inovi{\'c}}, {Laki{\'c}evi{\'c}}, and
  {Popovi{\'c}}]{kovacevicdojcinovicetal22}
{Kova{\v{c}}evi{\'c}-Doj{\v{c}}inovi{\'c}}, J.; {Doj{\v{c}}inovi{\'c}}, I.;
  {Laki{\'c}evi{\'c}}, M.; {Popovi{\'c}}, L.{\v{C}}.
\newblock {Tracing the outflow kinematics in Type 2 active galactic nuclei}.
\newblock {\em \aap} {\bf 2022}, {\em 659},~A130,
  \href{http://xxx.lanl.gov/abs/2112.11797}{{\normalfont
  [arXiv:astro-ph.GA/2112.11797]}}.
\newblock {\url{https://doi.org/10.1051/0004-6361/202141043}}.

\bibitem[{Collin} \em{et~al.}(2002){Collin}, {Boisson}, {Mouchet}, {Dumont},
  {Coup{\'e}}, {Porquet}, and {Rokaki}]{collinetal02}
{Collin}, S.; {Boisson}, C.; {Mouchet}, M.; {Dumont}, A.M.; {Coup{\'e}}, S.;
  {Porquet}, D.; {Rokaki}, E.
\newblock {Are quasars accreting at super-Eddington rates?}
\newblock {\em \aap} {\bf 2002}, {\em 388},~771--786,
  \href{http://xxx.lanl.gov/abs/astro-ph/0203439}{{\normalfont
  [arXiv:astro-ph/astro-ph/0203439]}}.
\newblock {\url{https://doi.org/10.1051/0004-6361:20020550}}.

\bibitem[{Li}(2012)]{li12}
{Li}, L.X.
\newblock {Accretion, growth of supermassive black holes, and feedback in
  galaxy mergers}.
\newblock {\em \mnras} {\bf 2012}, {\em 424},~1461--1470,
  \href{http://xxx.lanl.gov/abs/1205.0363}{{\normalfont
  [arXiv:astro-ph.CO/1205.0363]}}.
\newblock {\url{https://doi.org/10.1111/j.1365-2966.2012.21336.x}}.

\bibitem[{Bischetti} \em{et~al.}(2017){Bischetti}, {Piconcelli}, {Vietri},
  {Bongiorno}, {Fiore}, {Sani}, {Marconi}, {Duras}, {Zappacosta}, {Brusa},
  {Comastri}, {Cresci}, {Feruglio}, {Giallongo}, {La Franca}, {Mainieri},
  {Mannucci}, {Martocchia}, {Ricci}, {Schneider}, {Testa}, and
  {Vignali}]{bischettietal17}
{Bischetti}, M.; {Piconcelli}, E.; {Vietri}, G.; {Bongiorno}, A.; {Fiore}, F.;
  {Sani}, E.; {Marconi}, A.; {Duras}, F.; {Zappacosta}, L.; {Brusa}, M.;
  et~al.
\newblock {The WISSH quasars project. I. Powerful ionised outflows in
  hyper-luminous quasars}.
\newblock {\em \aap} {\bf 2017}, {\em 598},~A122,
  \href{http://xxx.lanl.gov/abs/1612.03728}{{\normalfont [1612.03728]}}.
\newblock {\url{https://doi.org/10.1051/0004-6361/201629301}}.

\bibitem[{Sanders} \em{et~al.}(1988){Sanders}, {Soifer}, {Elias}, {Madore},
  {Matthews}, {Neugebauer}, and {Scoville}]{sandersetal88}
{Sanders}, D.B.; {Soifer}, B.T.; {Elias}, J.H.; {Madore}, B.F.; {Matthews}, K.;
  {Neugebauer}, G.; {Scoville}, N.Z.
\newblock {Ultraluminous infrared galaxies and the origin of quasars}.
\newblock {\em \apj} {\bf 1988}, {\em 325},~74--91.
\newblock {\url{https://doi.org/10.1086/165983}}.

\bibitem[{Sanders} \em{et~al.}(2009){Sanders}, {Kartaltepe}, {Kewley}, {U},
  {Yuan}, {Evans}, {Armus}, and {Mazzarella}]{sandersetal09}
{Sanders}, D.B.; {Kartaltepe}, J.S.; {Kewley}, L.J.; {U}, V.; {Yuan}, T.;
  {Evans}, A.S.; {Armus}, L.; {Mazzarella}, J.M.
\newblock {Luminous Infrared Galaxies and the ``Starburst-AGN Connection''}.
\newblock In Proceedings of the The Starburst-AGN Connection; {Wang}, W.;
  {Yang}, Z.; {Luo}, Z.; {Chen}, Z., Eds.,  2009, Vol. 408, {\em Astronomical
  Society of the Pacific Conference Series}, p.~3.

\bibitem[{Collin} and {Zahn}(1999)]{collinzahn99}
{Collin}, S.; {Zahn}, J.P.
\newblock {Star formation and evolution in accretion disks around massive black
  holes.}
\newblock {\em A\&Ap} {\bf 1999}, {\em 344},~433--449.

\bibitem[{Cheng} and {Wang}(1999)]{chengwang99}
{Cheng}, K.S.; {Wang}, J.M.
\newblock {The Formation and Merger of Compact Objects in the Central Engine of
  Active Galactic Nuclei and Quasars: Gamma-Ray Burst and Gravitational
  Radiation}.
\newblock {\em \apj} {\bf 1999}, {\em 521},~502--508,
  \href{http://xxx.lanl.gov/abs/astro-ph/9908228}{{\normalfont
  [arXiv:astro-ph/astro-ph/9908228]}}.
\newblock {\url{https://doi.org/10.1086/307572}}.

\bibitem[{Lin}(1997)]{lin97}
{Lin}, D.N.C.
\newblock {Star/Disk Interaction in the Nuclei of Active Galaxies}.
\newblock In Proceedings of the IAU Colloq. 159: Emission Lines in Active
  Galaxies: New Methods and Techniques; {Peterson}, B.M.; {Cheng}, F.Z.;
  {Wilson}, A.S., Eds.,  1997, Vol. 113, {\em Astronomical Society of the
  Pacific Conference Series}, p.~64.

\bibitem[{Padovani} and {Matteucci}(1993)]{padovanimatteucci93}
{Padovani}, P.; {Matteucci}, F.
\newblock {Stellar Mass Loss in Elliptical Galaxies and the Fueling of Active
  Galactic Nuclei}.
\newblock {\em \apj} {\bf 1993}, {\em 416},~26.
\newblock {\url{https://doi.org/10.1086/173212}}.

\bibitem[{Zoccali} \em{et~al.}(2003){Zoccali}, {Renzini}, {Ortolani},
  {Greggio}, {Saviane}, {Cassisi}, {Rejkuba}, {Barbuy}, {Rich}, and
  {Bica}]{zoccalietal03}
{Zoccali}, M.; {Renzini}, A.; {Ortolani}, S.; {Greggio}, L.; {Saviane}, I.;
  {Cassisi}, S.; {Rejkuba}, M.; {Barbuy}, B.; {Rich}, R.M.; {Bica}, E.
\newblock {Age and metallicity distribution of the Galactic bulge from
  extensive optical and near-IR stellar photometry}.
\newblock {\em \aap} {\bf 2003}, {\em 399},~931--956,
  \href{http://xxx.lanl.gov/abs/astro-ph/0210660}{{\normalfont
  [arXiv:astro-ph/astro-ph/0210660]}}.
\newblock {\url{https://doi.org/10.1051/0004-6361:20021604}}.

\bibitem[{Gonzalez} and {Gadotti}(2016)]{gonzalezgadotti16}
{Gonzalez}, O.A.; {Gadotti}, D.
\newblock {The Milky Way Bulge: Observed Properties and a Comparison to
  External Galaxies}.
\newblock In Proceedings of the Galactic Bulges; {Laurikainen}, E.; {Peletier},
  R.; {Gadotti}, D., Eds.,  2016, Vol. 418, {\em Astrophysics and Space Science
  Library}, p. 199,  \href{http://xxx.lanl.gov/abs/1503.07252}{{\normalfont
  [arXiv:astro-ph.GA/1503.07252]}}.
\newblock {\url{https://doi.org/10.1007/978-3-319-19378-6_9}}.

\end{thebibliography}
\PublishersNote{}
\end{adjustwidth}
\end{document}